%% file: main.tex
\documentclass[conference,11pt]{IEEEtran}
\usepackage{url}
\usepackage{graphicx}
\usepackage{subfigure}
\usepackage{amssymb}
\usepackage{adjustbox}
\usepackage{array}
\usepackage{cite}
\usepackage{balance}
\usepackage{booktabs}
\usepackage{multirow}
\usepackage{pifont}

\newcolumntype{R}[2]{%
    >{\adjustbox{angle=#1,lap=\width-(#2)}\bgroup}%
    l%
    <{\egroup}%
}

\usepackage{graphicx}
\graphicspath{ {resources/} }

\makeatletter
\def\subsubsection{\@startsection{subsubsection}{3}{\parindent}{0ex plus 0.1ex minus 0.1ex}%
{0ex}{\normalfont\normalsize\itshape\textbf}}%
\def\paragraph{\@startsection{paragraph}{4}{2\parindent}{0ex plus 0.1ex minus 0.1ex}%
{0ex}{\normalfont\normalsize\itshape\textbf}}%
\makeatother

\usepackage{chronology}

\begin{document}

\date{}

\title{Amplification and DRDoS Attack Defense -- \\ A Survey and New Perspectives}

\author{
\IEEEauthorblockN{Fabrice J. Ryba\IEEEauthorrefmark{1}, Matthew Orlinski\IEEEauthorrefmark{1}, Matthias W\"ahlisch\IEEEauthorrefmark{1}, Christian Rossow\IEEEauthorrefmark{2}, Thomas C. Schmidt\IEEEauthorrefmark{3}}
\IEEEauthorblockA{\IEEEauthorrefmark{1} Freie Universit\"at Berlin, Berlin,
Germany, \\ Email: \{fabrice.ryba, m.orlinski, m.waehlisch\}@fu-berlin.de}	
\IEEEauthorblockA{\IEEEauthorrefmark{2} Saarland University, Saarbruecken,				
Germany, \\Email: crossow@mmci.uni-saarland.de}
\IEEEauthorblockA{\IEEEauthorrefmark{3} HAW Hamburg, Hamburg, Germany,
\\ Email: t.schmidt@haw-hamburg.de}	
}

\maketitle




\input{abstract}

\input{intro}
\input{drdos-background}
\input{DRDoS_Ana}
\input{IP_Spoofing_Detection}

\input{DRDoS_Detection}

\input{conclusion}

\balance
\bibliographystyle{IEEEtran}
\bibliography{DRDoS,ip-spoofing,related,rfcs}


\end{document}

%% file: abstract.tex
\begin{abstract}
The severity of amplification attacks has grown in recent years with recent attacks involving several hundreds of Gbps attack volume.
In this paper, we survey the threat of amplification attacks and compare a wide selection of different proposals for preventing, detecting, and filtering amplification attack traffic. 
Since source IP spoofing plays an important part in almost all of the amplification attacks surveyed, we also survey state-of-the-art of spoofing defenses as well as approaches to trace spoofing sources. 
This work acts as an introduction into many different types of amplification attacks and source IP address spoofing. By combining previous works into a single comprehensive discussion we hope to prevent redundant work and encourage others to find practical solutions for defending against future amplification attacks.
\end{abstract}

%% file: intro.tex
\section{Introduction}
\label{sec:intro}

Denial of Service (DoS) attacks against networks attempt to make a system or an entire network unavailable for its intended purpose~\cite{rfc4732}.
In some cases DoS attacks cause a complete loss of Internet connectivity for the victim~\cite{greennet}. The motivation for DoS attacks can be financial~\cite{l-usccf-04,prolexic}, political~\cite{greennet}, ideological~\cite{psmdb-aawpa-10}, reputation~\cite{cs-mhciw-08,PhantomL0rd}, to test new techniques and prepare for larger attacks, or purely for disruptive purposes~\cite{p-dtbi-13}. 



A general overview of DoS attacks and defense strategies can already be found in the surveys by Mirkovic and Reiher~\cite{mr-tdadd-04}, Douligeris and Mitrokotsa~\cite{dm-dadmc-04}, and more recently Zargar et al.~\cite{zjt-asdma-13}. Instead, this paper focuses on a particular type of DoS attack called an \emph{amplification attack} (also called Distributed Reflective Denial of Service (DRDoS) attacks~\cite{lwys-tda-11,r-ahrnp-14}) where the attacker seeks to maximise the volume of attack traffic sent to the victim whilst minimising the volume of traffic they send to trigger the attack~\cite{ve-dnsaa-06,kmgg-fsdaa-07,sasgm-mcdac-2009,akklg-daar-13,yy-pmcda-13,khrh-ehria-14,bbger-doaap-15}. We will not discuss other low traffic volume and high impact DoS attacks, e.g., Slowloris~\cite{h-s-09} or TCP SYN floods~\cite{rfc4987}, unless their defenses are also applicable to amplification attacks.
Generally, the adversary in amplification attacks targets vulnerabilities in Internet protocols and services to amplify the amount of attack traffic. 
The ease of performing amplification attacks and greater understanding of their effect has led to an increase of attacks in recent years. This is supported, e.g., by the OpenDNS Security Lab that saw over 5,000 different amplification attacks every hour in 2014~\cite{c-dnsamplification-2014}, and by Rossow's survey of amplification vulnerabilities~\cite{r-ahrnp-14}.

The most prominent form of amplification attack seen in recent years abuses the lack of endpoint verification in the Internet Protocol (IP) in order to trick third-party servers into sending large amounts of data to victims. That is, attackers spoof source IP addresses~\cite{b-sptps-89} to hide their identity and cause third-parties to send data to the victim as identified by the source address field of the IP packet. This is also sometimes called \emph{reflection} because attackers ``reflect'' attack traffic off of benign services.


As well as reflection, attackers sometimes strive to maximize the attack bandwidth sent to the victim by using \emph{amplification}. For example, many UDP-based protocols (such as DNS or NTP) that have a higher response payload size than the requests can be abused to amplify the reflected attack traffic. Consider DNS, while DNS lookup requests are typically rather small, the responses may be significantly larger, e.g., due to verbose information such as DNSSEC records.

By combining reflection and amplification attackers can generate attack traffic volumes which are significantly higher than their uplink bandwidth~\cite{greennet,prolexic,p-dtbi-13,p-tdbna-14}. Furthermore, not only single hosts may be affected by such attacks. Entire networks may struggle to cope with the extra bandwidth and processing demands~\cite{p-tdbna-14}.

Defending against this form of amplification attack is challenging for network operators.
Firstly, the attack traffic blends in benign communication and may be hard to distinguish from legitimate traffic.
Secondly, the attack packets received by victims are reflected by many distributed benign third-parties such as open DNS resolvers, effectively hiding the attacker's identity~\cite{openresolver}.

In this survey paper, we systematically structure existing research on amplification attacks to assist future research in this direction.
To the best of our knowledge, we are the first to categorize an research area that spans almost 200 publications and has become an important problem both in academia and practice.
To this end, we first compare various definitions and types of amplification attacks.
With this foundation, we then systematize \emph{preventive} countermeasures, and particularly look at the problem of (identifying and mitigating) IP spoofing.
We complete our survey with works that proposed \emph{reactive} countermeasures to defend against ongoing amplification attacks.
Our main contribution is thus a comprehensive survey of existing literature that until now was scattered.

\begin{figure}[t!]
\centering
  \includegraphics[width=0.475\textwidth]{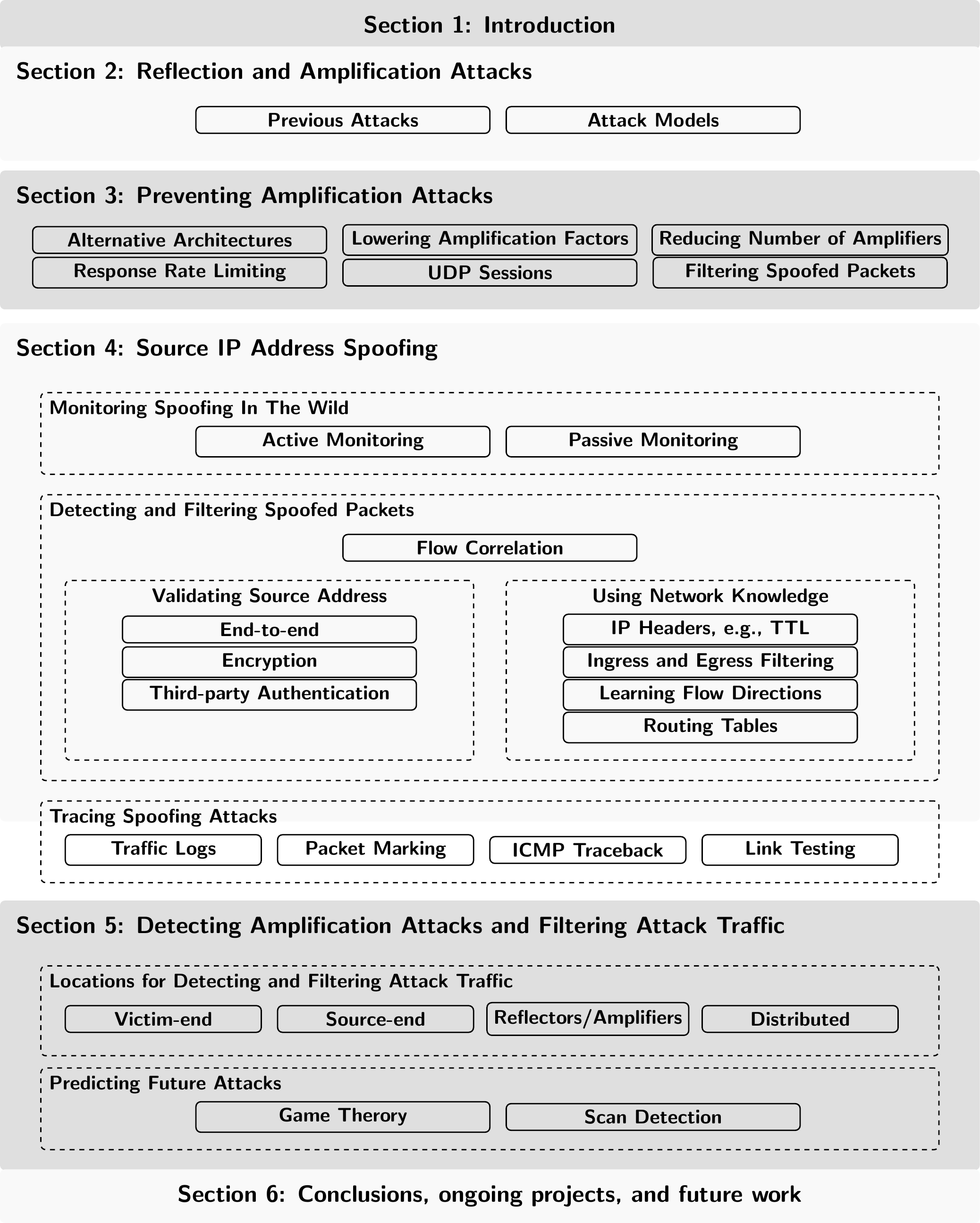}
  \caption{Categorisation of the amplification DDoS literature.}
  \label{fig:layout}
\end{figure}

The rest of this paper is organised as shown in Figure~\ref{fig:layout}. In Section~\ref{sec:amplOver}, we present prominent incidents from the past and discuss amplification attacks in great detail. We then describe methods to prevent amplification attacks in Section~\ref{sec:prevProt}. Then special attention is paid to preventing reflection and IP spoofing in Section~\ref{sec:ipSpoofing}. We finish the discussion about the solution space in Section~\ref{sec:reacProt} by presenting approaches to detect and filter amplification attacks. Finally, we conclude in Section~\ref{sec:conclusion}.

%% file: drdos-background.tex
\section{Amplification Attacks}
\label{sec:amplOver}

This section provides a detailed introduction into the different techniques for amplifying attack traffic. We start by giving a brief taxonomy and timeline of different types of amplificaiton attacks, and then we end the section by describing the techniques used in more detail.

\subsection{Taxonomy and Historical Background}
\label{sec:history}

\begin{figure*}
\centering
  \includegraphics[width=1\textwidth,natwidth=100,natheight=100]{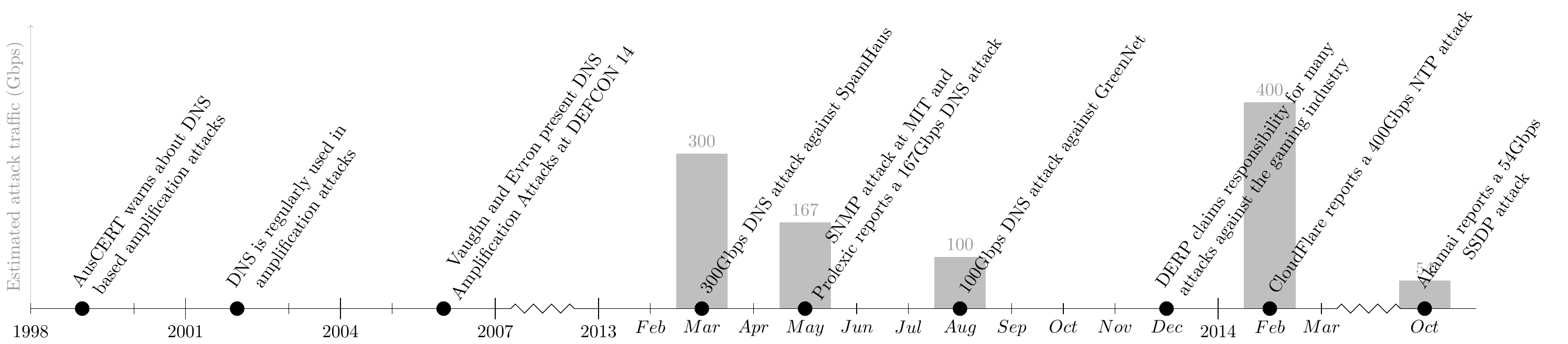}
\caption{Timeline of amplification attacks and related events.}
\label{fig:timeline}
\end{figure*}

There are many ways of amplifying the volume of attack traffic in DoS attacks. In this paper we give special focus to amplification attacks which use UDP based amplifying protocols and reflection. However, it is important to be able to distinguist between the different types of amplificaiton attacks. 

\subsubsection{DNS Servers as Amplifiers}

In 1999, AusCERT~\cite{a-dnsdos-99} warned about amplification attacks where attackers trick DNS servers into sending large amounts of data to spoofed IP addresses~\cite{b-sptps-89}. In 2000, the CIAC warned of Distributed Denial of Service (DDoS) attacks where attackers can send spoofed IP packets from many different locations~\cite{c-ddost-00}.

Today, DNS is widely used in amplification attacks because a 60 byte request from an attacker can easily generate a 512 byte response~\cite{ve-dnsaa-06,greennet,p-dtbi-13}. Furthermore, DNS servers which support Extension mechanisms for DNS (EDNS)~\cite{RFC-6891} can generate responses which are nearly 100 times larger than requests~\cite{r-ahrnp-14}. Some attackers even ensure large responses by creating domains specifically to use in amplification attacks~\cite{c-dnsamplification-2014}.

Any public DNS server which is configured to respond to requests for hosts outside of their domain can be used in amplification attacks. DNS servers that are configured to respond to all such requests are known as open DNS resolvers. According to the Open DNS Resolver Project approximately 25 million open DNS resolvers currently ``pose a significant threat'' to the Internet~\cite{openresolver}. It is also not difficult to find enough open DNS resolvers to use in amplification attacks. Rossow showed that it currently only takes 90 seconds to acquire a list of 100,000 open DNS resolvers~\cite{r-ahrnp-14}. 

\subsubsection{Other UDP Based Services} The DNS based amplification attacks just described use source IP address spoofing to trick third-parties into attacking a victim~\cite{b-sptps-89}. 

Source IP address spoofing also makes it very difficult for the victim to track attacker because all attack traffic they receive will come from third-parties. UDP is also used by attackers as the transport layer protocol because it is stateless and has no built in mechanism to verifying the source IP address of packets. 

Recently, another UDP based protocol (SSDP) has been used with source IP address spoofing in amplification attacks. The United States Computer Emergency Readiness Team (US-CERT) first issued a warning about SSDP in January 2014~\cite{us-udpba-14}, and in October 2014 it was used to generate 54Gbps of traffic in a single attack~\cite{a-ssdpr-14}. 

Other well documented DoS attacks that used UDP and source IP address spoofing are the MIT SNMP based attack~\cite{MITdrdos}, and the CloudFlare attack which reached a peak of nearly 400Gbps using 4,529 NTP servers~\cite{p-tdbna-14}. The huge volume of traffic generated in the CloudFlare attack was caused by a command in NTP called ``monlist'', which returns a list of the previous 600 IP addresses to access the server. Monlist does not affect the core functionality of NTP and should be disabled by upgrading to version 4.2.7 and above~\cite{us-ntpaa-14}.

\begin{table*}
\caption{Recent large scale amplification attacks and their properties}
\label{table:attacks}
\begin{tabular}{ llllm{3cm}m{5cm} }
\toprule
\textbf{Reference} & \textbf{Year} & \textbf{Amplification Protocol} & \textbf{Peak Traffic} & \textbf{Attacker} & \textbf{Description} \\ 
\midrule
MIT~\cite{MITdrdos} & 2013 & SNMP & Unknown & Unknown & Spoofed requests were directed to SNMP enabled devices on the MIT network which attacked devices outside the network. \\
\midrule
GreenNet~\cite{greennet} & 2013 & DNS & 100 Gbps & Unknown & Attackers brought down GreenNet, a hosting company who's customers including Privacy International and Zimbabwe Human Rights Forum. \\
\midrule
Prolexic~\cite{prolexic} & 2013 & DNS & 167Gbps & Unknown & The attack targeted a real-time financial exchange platform. \\
\midrule
Spamhaus~\cite{p-dtbi-13} & 2013 & DNS & 300Gbps & Cyberbunk & Over 30,000 DNS servers were involved in the amplification attack. Attackers also used SYN floods and prefix hijacking to disrupt the Internet connectivity of Spamhaus and Project Honeypot. \\
\midrule
PhantomL0rd~\cite{PhantomL0rd} & 2013 & NTP & 100Gbps & A hacker group named DERP claimed responsibility on Twitter. & Attackers targeted game servers and streaming sites. Attacks against the games industry are becoming increasingly popular~\cite{a-tsofi-14}.\\
\midrule
CloudFlare~\cite{p-tdbna-14} & 2014 & NTP & 400Gbps & Unknown & Latency increased all over Europe following attack using 4,529 NTP servers. French hosting firm OVH also claimed to have seen an attack over 350Gbps at the same time.\\
\midrule
Akamai~\cite{a-ssdpr-14}& 2014 & SSDP & 54Gbps & Unknown & Many of the devices used in this attack were home-based UPnP devices.\\
\bottomrule
\end{tabular}
\end{table*}

\subsubsection{Smurf and Fraggle Attacks}

As well as misusing UDP based services, there are many other ways of amplifying attack traffic volume. For example, prior to 1998 attackers were sending ICMP requests to the broadcast address of networks whilst spoofing the source IP address of victims. The routers receiving the ICMP requests broadcast the requests to all of the hosts on their network. These attacks are known as Smurf attacks~\cite{cert-smipd-98, ks-sbddos-07}, and in Smurf attacks the routers which broadcast the requests act as amplifiers because they increase the volume of attack traffic. 

End hosts which respond to requests after they are broadcast may be amplifiers as well. For example, in Fraggle attacks an attacker may send CharGen requests to the broadcast address of many networks at the same time~\cite{hmowt-mtdos-01}. End hosts which receive CharGen requests may respond with much more data than they receive, thus the end hosts in Fraggle attacks are amplifiers as well as the routers which broadcast the request. 

\subsubsection{Fork Loops} 

There is also another category of amplification attacks which we have not yet covered. In 2009 Shankesi et al.~\cite{s-aavfp-06} described an amplification attack as being where "the number of messages on the network can amplify to essentially an arbitrary large number"~\cite{sasgm-mcdac-2009}. It is important to note however that this definition doesn't take into account the size difference between request and response packets, which is an important factor in the UDP based amplification attacks we have already mentioned. 

The amplification attacks described by Shankesi et al. are caused by ``fork loops'' which are used to attack VoIP networks running SIP~\cite{s-aavfp-06}. Fork loops are situations where requests are sent between SIP proxies indefinitely and at least one extra request is generated every iteration~\cite{sasgm-mcdac-2009}. Using this type of amplification attack it is possible to cause 2 SIP proxies to exchange up to $2^{70}$ duplicate requests consuming their resources and preventing them from responding to legitimate requests~\cite{s-aavfp-06}.

\subsection{Responding to a Growing Threat}

Today, Smurf, Fraggle, and Fork Loop attacks have largely been countered by better network configuration and patches to SIP respectively. However, attacks involving reflection and UDP based amplification are growing in number. 

Akamai detected 9 DoS attacks involving NTP, CharGen, and SSDP which peaked over 100 Gbps in the last 3 months of 2014. This is three times more than in the same period in 2013~\cite{a-tsofi-14}. 

Figure~\ref{fig:timeline} and Table~\ref{table:attacks} illustrate that the bandwidth used in amplification attacks is also growing. In March 2013 an amplification attack was launched against Spamhaus \cite{p-dtbi-13}. The DNS based attack reached an estimated peak of about 300Gbps  and was the biggest DoS attack ever recorded~\cite{b-pceobc-13}. Nearly a year later and an even bigger attack reportedly reached a peak of 400Gbps by using NTP~\cite{p-tdbna-14}. 

\label{sec:anycast}
The attack traffic generated during the Spamhaus attack came from over 30,000 different DNS servers. In order to guard against this and future attacks, they deployed anycast routing and a distributed Content Delivery Network (CDN)~\cite{p-dtbi-13}. Anycast involves BGP routers simultaneously advertising the same destination IP address so that attack traffic can be absorbed by many different data centers. However, relying on anycast and CDNs may be a short term solution, and it can only prevent amplification attacks when amplifiers and anycast routes are evenly distributed around the Internet.

The CloudFlare attack came 1 month after a ``monlist'' warning issued by US-CERT~\cite{us-ntpaa-14}, and during a 13-week period in which K{\"u}hrer et al. reported a 92\% drop in the number of vulnerable NTP servers~\cite{khrh-ehria-14}. As a consequence we may need to reevaluate how we respond to threats as they are discovered. For example, alerting system administrators about protocol vulnerabilities may prompt attackers to abuse the identified amplification vectors.

\subsection{Problem Description}
\label{sub:problemdescription}

In this section we will provide high level descriptions of the attacks and concepts seen so far. We will start by discussing the amplification factor of different attacks and go on to describe the building blocks of the most common amplification attacks.

\subsubsection{Amplification Factor}

The amplification factor of an attack ($X(P)$) for a protocol ($P$) is the ratio of the total number of bytes in the amplified traffic (explicitly including headers, padding and payloads of all protocols involved, such as Ethernet, IP, UDP or the application-level protocol) and the number of bytes sent by the attacker (same calculation), as shown in Equation~\ref{eq:factor}. 

\begin{equation} \label{eq:factor}
 X(P) = \frac{number\,of\,response\,bytes(P)}{number\,of\,request\,bytes(P)} 
\end{equation}

Our definition is slightly different from the one used in \cite{r-ahrnp-14}, as we explicitly include all lower-level protocols, whereas Rossow focused on the number of UDP payload bytes. Still, the amplification factors provided by Rossow~\cite{r-ahrnp-14} offer a convenient way of comparing protocols and can help with prioritization. For example, Rossow found that NTP had the highest amplification factor all of the protocols he tested. He found that if attacks carefully select the NTP servers to maximize their attack they can achieve an amplification factor of 4670, which when compared to 64 for open DNS resolvers shows us how urgent it is to update existing NTP infrastructure.


\subsubsection{Reflection}
\label{sec:reflection}

\begin{figure}[h]
\centering
\includegraphics[width=0.4\textwidth,natwidth=100,natheight=100]{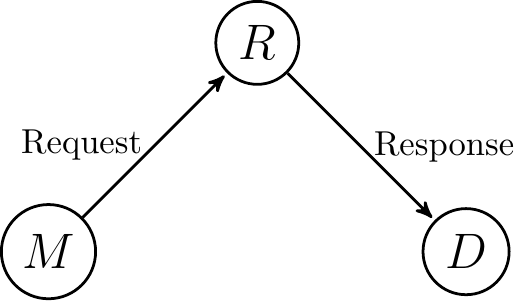}
\caption{Reflection attack. An attacker ($M$) sends a spoofed request to a reflector ($R$) which responds to the final destination ($D$).}
\label{fig:reflection}
\end{figure}

Many of the amplification attacks in the past were possible because the attacker can spoof the address of the victim causing traffic to be ``reflected'' by third-party servers, e.g., by open DNS resolvers~\cite{rkm-dadra-13} (see Section~\ref{sec:history}). Figure~\ref{fig:reflection} shows a reflection attack where an attacker ($M$) changes the source address in the request it sends to the reflector ($R$). Not knowing that the source address was spoofed, the reflector sends its response to the final destination ($D$, i.e., the victim).

The key advantages of reflection attacks for the attacker are: (a) attackers hide their identity from $D$ because all traffic received by $D$ comes from third-parties; (b) attackers can trigger attacks coming from different geographic or topological regions of the Internet; and (c) attackers do not receive responses and therefore do not risk using up their available download bandwidth.

One example of a DoS attack which uses reflection is a TCP SYN/ACK attack, where an attacker sends spoofed SYN requests to a reflector so that the reflector sends SYN/ACK packets to the victim~\cite{p-aurdd-01}. 

As well as the final destination as indicated by the spoofed source IP address, it is also possible for reflectors and those that use them legitimately to be victims of reflection attacks. For example, in SYN flooding attacks~\cite{c-dafdd-02,cy-datsf-06,njh-dsfau-08,nh-bspsm-09,kc-mmtsf-11,gh-ldais-12}, the attacker floods a reflector with spoofed SYN requests and changes the source IP address of packets so they do not receive replies. The aim of SYN flooding attacks is not to attack the final destination indicated by the spoofed IP address, but to consume enough resources at the reflector to make it unresponsive to legitimate traffic. The adversary can also attack multi-homed networks using reflectors. For example, if the reflector is connected to the Internet by at least two uplinks, and the link between the reflector and the final destination offers significantly lower bandwidth compared to the link between the attacker and reflector. Then by carefully selecting a spoofed source address which is reachable via the low bandwidth connection, the attacker can implement a DoS attack on this link.

It is worth mentioning again, however, that not all of the amplification attacks described in this paper use reflection, e.g., fork loop based attacks do not~\cite{sasgm-mcdac-2009}.


\subsubsection{Amplification}
\label{sec:amplification}

\begin{figure}[h!]
\centering
\includegraphics[width=0.25\textwidth,natwidth=100,natheight=100]{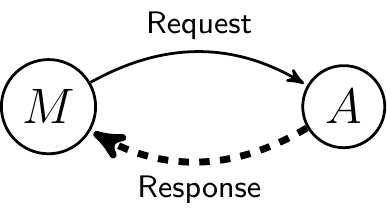}
\caption{Simple amplification attack. An attacker ($M$) sends a request to an amplifier ($A$) which responds with more data than it received. The response line is dashed in this instance to indicate that the attacker may not receive the response.}
\label{fig:ampwithoutIP}
\end{figure}

A key factor in amplification attacks is the ability of attackers to trigger responses which are significantly larger than the requests, either in terms of bytes or number of packets. Figure~\ref{fig:ampwithoutIP} describes the basic scenario without reflection where an attacker sends a request to an amplifier ($A$) which responds with a higher volume of data than it receives. In this scenario the amplifier, or the network between the attacker and the amplifier, are the victim because they cannot cope with the amplified traffic.
The main advantage of such an attack is twofold.
First, the adversary saves resources with respect to its network uplink.
This is important in case of mobile scenarios.
Second, discovering the trigger of the attack is more challenging.
In contrast to high volume flooding attacks, amplification triggers are hardly visible on frequently used uplinks, which complicates early detection.

It is worth noting that the attacker model implies that the attacker is either capable to handle potential response traffic or that the attacker will not receive the response data.
The latter follows directly in case of a successful attack but can also be achieved by address renewals at the attacker side.

Pure amplification attacks will be of increased importance with the upcoming deployment of the Internet of Things (IoT).
IoT networks are characterized by constrained devices with limited network capacity, typically connected via border routers to the Internet.
Congesting the path from some of the IoT devices to the gateway may lead to disconnecting the complete IoT domain.


\subsubsection{Reflection and Amplification Together}

\begin{figure}[h!]
\centering
  \includegraphics[width=0.5\textwidth,natwidth=100,natheight=100]{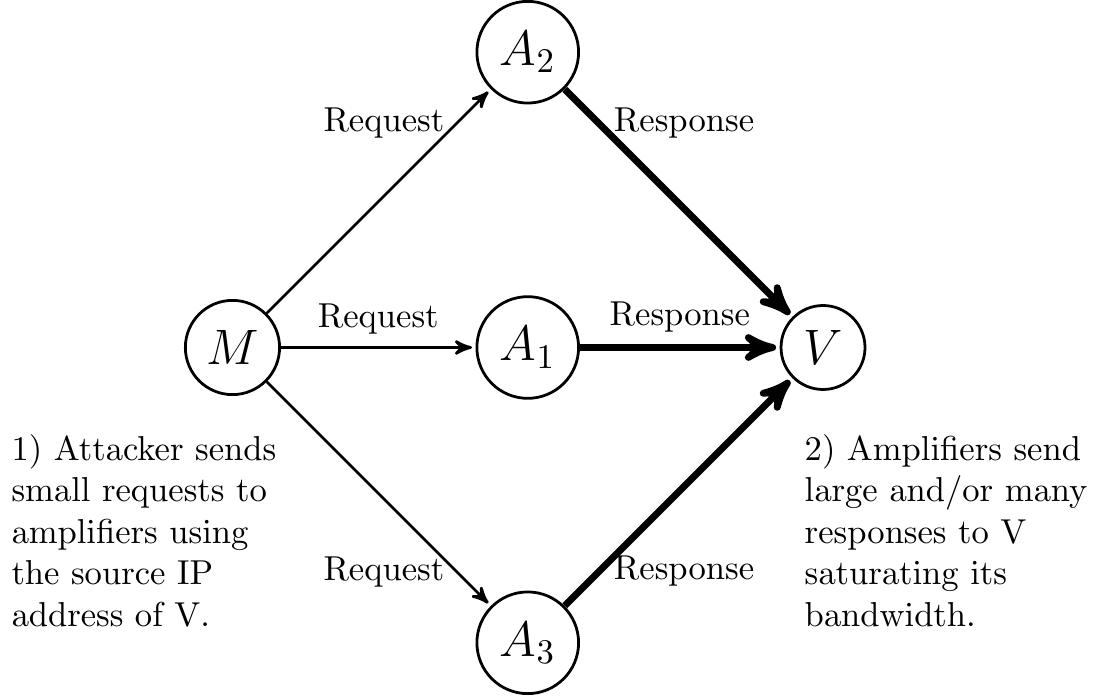}
  \caption{An attacker ($M$) sends spoofed requests to many amplifiers ($A$) causing amplified responses to be sent to the victim ($V$).}
  \label{fig:drdos}
\end{figure}

All of the amplification attacks seen in Table~\ref{table:attacks} combine reflection and amplification, a high level overview of these attacks is given in Figure~\ref{fig:drdos}. In this model an attacker ($M$) sends requests using the amplification protocol ($P$) to the amplifiers ($A$) but also uses reflection so that the amplifiers respond to the victim ($V$). 

The volume of attack traffic which can be sent to $V$ depends on the accumulated available bandwidth between $M$ and $A_1 ... A_n$, the accumulated available bandwidth between $A_1 ... A_n$ and $V$, and the amplification factor of $P$. Depending on the success of the attack, the volume of traffic sent to the victim can result in a complete loss of connectivity for the victim and their network. 

\subsubsection{Amplification Attacks Using a Botnet}
Most amplification attacks are launched from single origins~\cite{kkmnk-amdad-15} and just appear to be distributed due to the large set of amplifiers.
However, a single attacker may not have sufficient uplink bandwidth to send requests to many amplifiers at the same time. Attackers can target more amplifiers, and further obscure the origin of attacks, by means of a botnet~\cite{fsr-sbbd-09,akklg-daar-13}. 


The owners of bot-infected systems are unaware that their machines are being used in DDoS attacks. Alternatively, in crowd-sourced DDoS botnets, participants willingly install attacking software on their machines, e.g., the Low Orbit Ion Cannon (LOIC) software~\cite{p-loicn-}.
Such volunteer botnets were used in \emph{Operation Payback} to attack global companies, such as MasterCard, Visa, and PayPal~\cite{psmdb-aawpa-10}.

\begin{figure}[h!]
\centering
  \includegraphics[width=0.5\textwidth,natwidth=100,natheight=100]{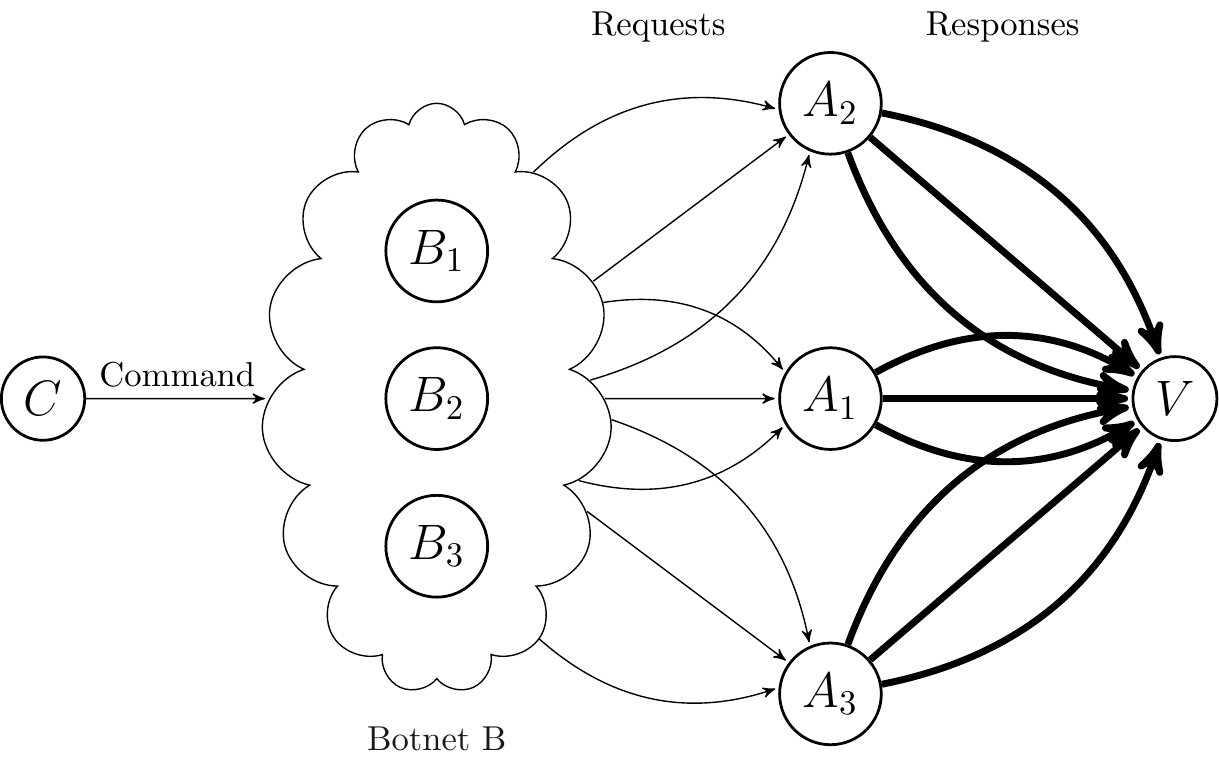}
  \caption{An attacker uses a controller ($C$) to instruct a botnet ($B$) to simultaneously send spoofed requests to many amplifiers ($A$), causing amplified responses to be sent to the victim ($V$).}
  \label{fig:botnets}
\end{figure}

Figure~\ref{fig:botnets} shows one model of an amplification attack using a botnet. In this example the attacker uses a controller ($C$) to send instructions to the botnet ($B$). The bots then send spoofed requests to amplifiers at similar times, such that the bandwidth of $B_1...B_n$ accumulates. 


\subsubsection{Fraggle and Smurf Attacks}

\begin{figure}[h!]
\centering
  \includegraphics[width=0.5\textwidth,natwidth=100,natheight=100]{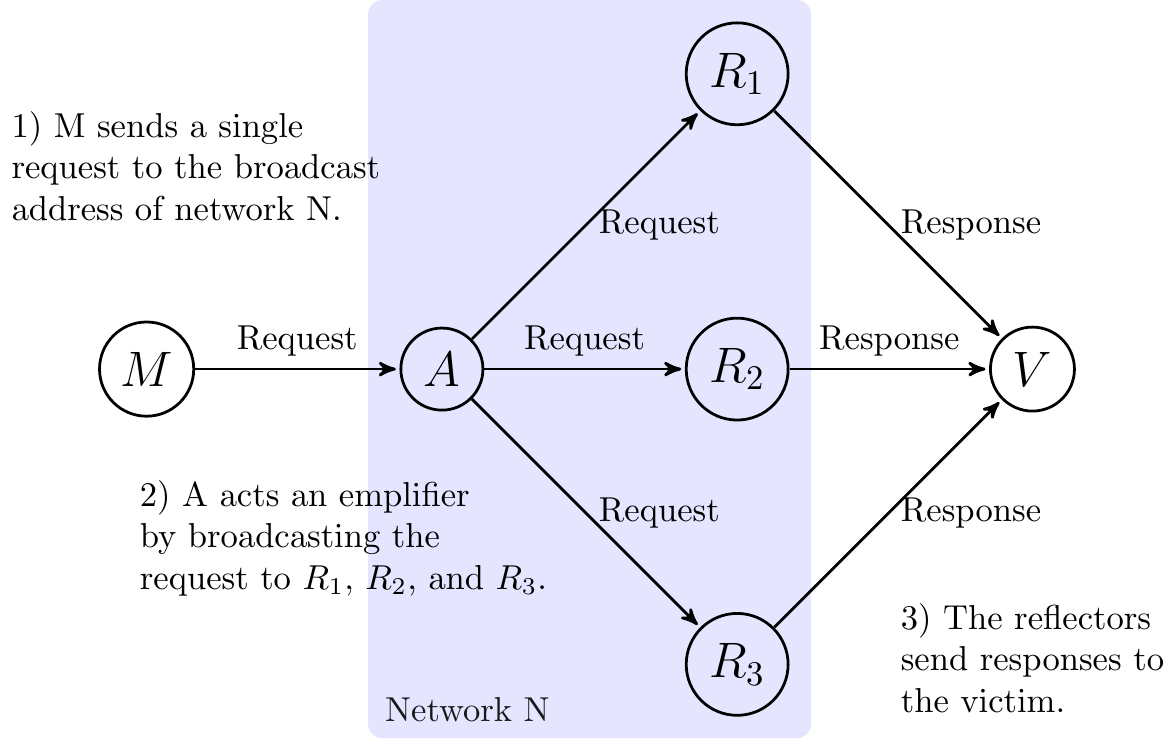}
  \caption{Smurf attack. An attacker ($M$) sends a spoofed request to the broadcast address of network ($N$). The amplifier ($A$) broadcasts the request to many reflectors ($R$) which respond to the victim ($V$).}
  \label{fig:smurfAttack}
\end{figure}

Fraggle and Smurf also use reflection and amplification, but their attacks look slightly different from the model in Figures~\ref{fig:drdos} and~\ref{fig:botnets}. Figure~\ref{fig:smurfAttack} shows a Smurf attack where $M$ sends a request to the broadcast address of network ($N$), $A$ acts as the amplifier by duplicating the request, and the reflectors ($R$) send responses to the victim $V$. Fraggle attacks are similar to Smurf attacks but the requests are amplified even further by the end hosts that receive them.

\subsubsection{Fork Loops}

Fork loops are different to the reflection and amplification attacks because they rely on messages recursively being sent between SIP proxies~\cite{sasgm-mcdac-2009}. Yet fork loops have also been referred to as amplification attacks because they provide a way for an attacker to amplify the number of requests they send to SIP proxies. 

\begin{figure}[h!]
\centering
  \includegraphics[width=0.5\textwidth,natwidth=100,natheight=100]{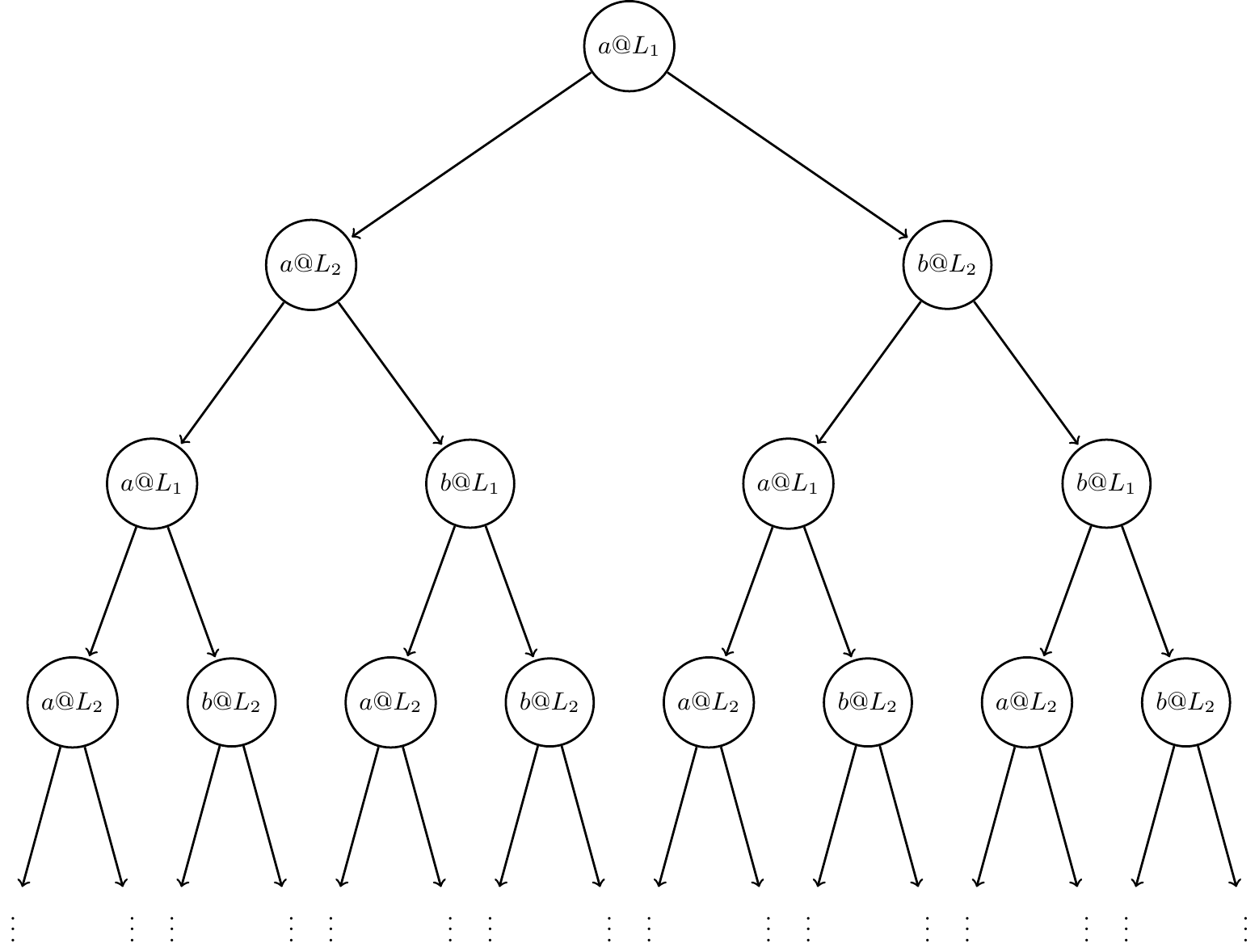}
  \caption{Fork Loops. In this example taken from \cite{s-aavfp-06} a single request spawns $2^n$ duplicates where $n$ is the number of iterations.}
  \label{fig:forkloop}
\end{figure}

Figure~\ref{fig:forkloop} shows one example of what can happen when two SIP proxies ($L_1$ and $L_2$) are not configured correctly and do not detect loops. It depicts the scenario described in \cite{s-aavfp-06} where a request for the user $a@L_1$ is sent to the proxy $L_1$. In this example, $L_1$ has two options for reaching $a@L_1$ which are both located at $L_2$. This causes $L_1$ to fork the request and send two copies to $a@L_2$ and $b@L_2$ which also both fork the request to $a@L_1$ and $b@L_1$. The scenario is caused by misconfiguration and can potentially result in $2^n$ duplicate requests created where $n$ is the number of iterations.

\subsection{Discussion}

In this section we have described many different types of amplification attacks. It is clear from our descriptions that the attacks seen in Table~\ref{table:attacks} are very different from Fraggle and Fork Loop attacks.

Therefore, this survey will focus on the types of amplification attacks seen in Table~\ref{table:attacks} and as described in Figures~\ref{fig:drdos} and~\ref{fig:botnets}. The major advantages of amplification attacks from an adversarial point of view are:

\begin{itemize}
  \item Attack traffic can be amplified beyond the attacker's upload bandwidth.
  \item The attacker remains anonymous by using reflection (IP address spoofing).
  \item The attack traffic at reflectors and amplifiers can be very difficult to distinguish from legitimate traffic.
  \item The attack complexity is rather low and it is thus relatively straightforward to get started with amplification attacks even for the wide audience.
\end{itemize}

%% file: DRDoS_Ana.tex
\section{Preventing Amplification Attack}
\label{sec:prevProt}
In this section, we will describe preventative measures for the amplification attacks. For example, we discuss preventative measures which amplifiers can use before responding to requests (e.g. response rate limiting), and more radical approaches such as alternative content delivery systems~\cite{jstpb-nnc-2009}.
We will discuss non-preventive (i.e., reactive) countermeasures, such as detection and filtering, in Section~\ref{sec:reacProt}.

\label{sec:prevAmpli}

\subsection{Alternative Internet Architectures}

Alternative Internet architectures such as Capability Based Architectures (CBA)~\cite{arw-pidc-04,ywa-tvadl-08} or Content-Centric Networking (CCN)~\cite{jstpb-nnc-2009} may remove existing DoS vectors. For example, in CBA, senders must first obtain ``permission to send'' from the destination before sending packets. Anderson et al.~\cite{arw-pidc-04} proposed that verification points around the network ensure that the source has permission to send data to the destination. To obtain permission, Anderson et al.~suggested that the source uses Request-To-Send (RTS) packets, which are routed between RTS servers to ensure the victim is not flooded.

However, the problem with the CBA system from Anderson et al.~\cite{arw-pidc-04}, and the more recent TVA~\cite{ywa-tvadl-08}, is that every packet must include some kind of extra ``capability'' information. Maintaining, exchanging, and checking capabilities creates a lot of extra overhead for networks, and the increased packet size may also increase fragmentation. 

CCN will remove the possibility of DNS based amplification attacks simply because DNS is no longer required. In CCN, ISPs can deploy content routers to cache data which many users request. However, CCN uses the request/response paradigm in the form of ``Interest'' and ``Data'' packets, and it is currently unknown how resilient CCN would be to amplification attacks or new forms of DoS attacks.

\subsection{Lowering Amplification Factors} 
\label{sec:lowerfactors}

Alternative, to remain compatible with the current Internet design, is fixing the protocols that are vulnerable to amplification.
One such approach is to lower the amplification factors, e.g., by increasing the size of the requests in vulnerable protocols. This would reduce the amplification factor, but also increase the traffic across the Internet, which is not desirable. Moreover, it is not reasonable to suggest that all protocols should be amplification free. For example, it is quite reasonable to expect that request packets for large files are smaller or fewer than the response.

A more practical approach for lowering the amplification factors may be to disable the redundant services that amplifiers offer. For example, Rossow showed that the Network Time Protocol (NTP) has the highest amplification factor out of the 14 UDP protocols he tested~\cite{r-ahrnp-14}. K\"uhrer et al.~achieved a reduction of the amplification factor of NTP by disabling the NTP services with the highest request/response factors~\cite{khrh-ehria-14}. Since the services with the highest request/response factors do not affect the core capabilities of NTP, these extra services can be disabled while still enabling time synchronization. 

Next to NTP, many other protocols (e.g. DNS, CharGen, SNMP) can be used for amplification attacks. Some DNS providers have recently started deprecating the DNS ANY request in response to the growing threat of amplification attacks. One provider, CloudFlare, now responds with the RCODE 4 ``Not Implemented''%
\footnote{\url{https://blog.cloudflare.com/deprecating-dns-any-meta-query-type/}}. However, identifying and disabling services with high request/response factors takes time, and may result in the loss of functionality for benign applications.

%
One technique to reveal protocols which suffer from amplification attacks due to misconfiguration or bad design can be model checking~\cite{sasgm-mcdac-2009}. By modeling the protocol using rewriting logic~\cite{m-rlmws-2000}, a set of system states can be generated and checked for amplification by comparing the cost of servicing requests against a predefined threshold%
\footnote{The cost in \cite{sasgm-mcdac-2009} is the number of duplicate packets generated by SIP proxies}. This approach can be used to detect flaws in protocols where the amplification factor at a single point might not seem significantly high, and it has been used to describe forking loops in SIP~\cite{sasgm-mcdac-2009}.


\subsection{Reducing the Number of Amplifiers}

Many amplifiers were used in the larger amplification attacks. For example, over 30,000 separate open DNS resolvers were used in the Spamhaus attack (See Table~\ref{table:attacks}). A number of projects listed in Section~\ref{sec:conclusion} are attempting to bring misconfigured amplifiers to the attention of system administrators so that they can be patched or configured correctly. However, this relies on widespread cooperation between administrators who may not have the knowledge or the resources to update their servers.


Another interesting point to make here is does the Internet need so many public services such as DNS servers? Perhaps more cooperation between network providers can lower the impact of amplification attacks by lowering the number of amplifiers required to operate the Internet.


\subsection{Response Rate Limiting}

Moreover, amplifiers could also be configured to respond to a limited number of request from each IP or network address within a given time frame. For example, support for response rate limiting is easily applied to DNS when using newer versions of BIND. However, rate response rate limiting protects against abuse of a single amplifier and an attacker may simply choose to use multiple amplifiers at a low request rate. It only takes an attacker a short amount of time to look for a sufficient number of different amplifiers (less than a minute in some case~\cite{r-ahrnp-14}). 

\subsection{Sessions for UDP}

Many amplification attacks rely on stateless communication via the UDP protocol in order to send large amounts of data to spoofed IP addresses. Those amplification attacks are made more difficult if UDP-based protocols required sessions to be opened (similar to the TCP three-way handshake) before large amounts of response data can be transmitted.
To counter this, one can include session information in UDP packets.
For example, clients using the Steam query protocol have to request a 4-byte challenge before they can request large amounts of information about a game server~\cite{v-smsqp-15}. The client has to append the challenge response to future requests. However, attackers may be able to use responses from session initialisation exchanges for amplification attacks so long as they can open new sessions for the same victim with many amplifiers, and can also eavesdrop on the victims Internet traffic to see responses.


A downside to session hardening UDP protocols is that it increases the packet sizes (and, depending on the implementation, requires additional packets for a handshake), which conflicts in particular in resource-constraint and real-time contexts. It may also add latency to the initial request to open a sessions. The most challenging aspect of session hardening existing UDP-based services is compatibility with existing clients. In order to prevent amplification attacks session support would have to be universal, and upgrading may cause problems with legacy systems.

\subsection{Differentiating Bots From Humans}

Also Completely Automated Public Turing tests to tell Computers and Humans Apart (CAPTCHAs)~\cite{kkjb-bsoda-05} would prevent attackers from using certain services in DoS attacks.
While successful to defend against application-level DoS attacks, CAPTCHAs are not successfully used in practice against amplification attacks, and we name them for completeness only.
Consider, for example, DNS, where lookups are often caused by a user visiting a website. If such DNS requests required that a user completes a CAPTCHA, then it would be harder for attackers to use open DNS resolvers for attacks. However, this may delay a users access to information and complicate legitimate automated services which use DNS. With this in mind, Oikonomou~\cite{ok-mhbda-09} attempted to model human behavior so that servers can automatically distinguish between requests from bots and requests from humans.

There is another problem with only replying to requests from humans, and that is that many legitimate automated processes need to make requests for information over the Internet. These processes are prevented from accessing services secured by a CAPTCHA or other bot detection mechanisms.

\subsection{Filtering Spoofed Packets} 
In most amplification attacks the attacker sends requests with the spoofed IP address of the victim. Filtering spoofed packets is considered one of the most effective countermeasures against amplification and other DoS attacks~\cite{bb-spies-05, msbvs-ida-06, m-sasdv-13, khrh-ehria-14}. However, in 2005 the MIT ANA Spoofer Project showed that around 25\% of ASes allowed spoofed IP packets to be sent out of their network, and the ability of attackers to launch amplification attacks shows that it is still a problem today. Recent estimates state that it is still possible to send spoofed IP packets from about 20\% of the Internet~\cite{bb-spies-05}. 
The next section will discuss spoofing in more detail and discuss some of the methods proposed to defend against it.

%% file: IP_Spoofing_Detection.tex
\section{Source IP Address Spoofing}
\label{sec:ipSpoofing}
An inherent requirement for amplification attacks in practice is IP address spoofing.
UDP and IP have no built-in mechanism to determine if the source address is spoofed, so amplifiers reply to the spoofed address instead of the original sender.
Using spoofing for malicious purposes was first discussed in 1989~\cite{b-sptps-89}, and a detailed analyses of the problem was carried out by Heberlein and Bishop in 1996~\cite{hb-acas-96}, and again by Dunigan in 2001~\cite{d-bsp-00}. Since then there have been many other surveys related to spoofing which we have summarized in Table~\ref{table:surveys}. These surveys can be broadly categorized as either focusing on spoofed packet detection/filtering~\cite{sou-isdi-09,el-sisd-09}, or tracing the attacker~\cite{ba-oit-03,js-dstm-09,vr-sitmo-10,jm-spml-13}.

Chen and Yeung~\cite{cy-datsf-06} also define three common IP spoofing techniques which all have different purposes, random spoofing, subnet spoofing, and fixed spoofing. In \emph{random spoofing}, the attacker randomly generates source addresses for the attacking packets. This technique is often used in TCP SYN flood attacks~\cite{rfc4987} where the destination of replies are not relevant to the success of the attack but the attacker still wants to obscure their identity. In \emph{subnet spoofing} the attacker will choose a source address in its own subnet, which can help to avoid some countermeasures such as ingress filtering, but limits the number of possible victims. \emph{Fixed spoofing} is the form of spoofing which is most commonly used in amplification attacks. This is where the attacker chooses the IP address of a single victim to be the source address on the spoofed packets, meaning that replies will be sent to the victim and not the attacker(s). 

\begin{table}
\begin{center}
\caption{Survey papers for filtering spoofed IP packets and detecting attackers.}
\label{table:surveys}
\begin{tabular}{ p{13.0em}p{12.0em} }
\toprule
\textbf{Survey Name} & \textbf{Papers Covered} \\ 
\midrule
Detecting Spoofed Packets~\cite{tl-dsp-03} & \cite{cnwvb-ddsin-99, swka-pnsit-00, sds-iidr-2000} \\ 
\midrule
IP Spoofing Defense: An Introduction~\cite{sou-isdi-09} & \cite{rfc2827, mjr-pisdt-06, bl-spm-05, pl-erpfd-01, dyc-cipti-08, onsw-peidf-07, lmewr-lvidi-08, sgd-pfsv-08, lkhp-bidmv-07, yps-ppimd-03, cs-urpfe-05, ls-tspbh-07, wjs-dasit-07, yps-snpmf-06, ga-prims-07, cpm-dstdd-07} \\ 
\midrule
On the State of IP Spoofing Defense~\cite{el-sisd-09} & \cite{wrl-savap-07, dfs-aait-02, t-arpof-03, a-tppmi-05, f-rosd-06, bl-spm-05, pl-erpfd-01, rfc4301, b-rcpf-04, t-arvpo-03,  b-sc-96, ffl-dinp-05, jws-hfeda-03, wjs-dasit-07, tl-dsp-03, rfc2827, rfc3013, rfc1812, rfc3704, llyxw-psasa-08, lmwrz-ssave-02, dyc-cipfc-06, lkhp-bidmv-07, yps-ppimd-03, yps-snpmf-06, el-idplv-07} \\ 
\midrule 
DDoS: Survey of Traceback Methods~\cite{js-dstm-09} & \cite{s-ciont-00, bc-tapta-00, bl-itm-00, swka-nsit-01, sp-aamsi-01, swka-pnsit-00, bn-afdpm-03, dfs-aait-02, spsjt-sit-02, xzg-fdpmi-09, pl-erpfd-01, zg-ttspa-06, mbhn-diuat-02, adhc-tbpmi-02} \\ 
\midrule
A Survey of IP Traceback Mechanisms to Overcome Denial-of-service Attacks~\cite{vr-sitmo-10} & \cite{gk-mpasi-08, swka-nsit-01, dkk-ursis-00, spsjt-sit-02, ag-nhsep-06, spslj-hit-01}  \\ 
\midrule
A Survey on Packet Marking and Logging~\cite{jm-spml-13} & \cite{gk-mpasi-08, ywsy-ppitt-12, spsjt-sit-02, swka-nsit-01, sp-aamsi-01, ag-nhsep-06, zg-ttspa-06, ba-itdpm-03, gs-tppma-07, g-ppmli-08, ww-tdpmi-10, dfs-aait-02, pl-eppmi-01, bm-tnats-02, spslj-hit-01} \\ 
\midrule
On IP Traceback~\cite{ba-oit-03} & \cite{cnwvb-ddsin-99,bl-itm-00,s-ciont-00,bc-tapta-00,mmwwz-deidi-01,swka-nsit-01,sp-aamsi-01,c-dafdd-02,ls-caat-02,lmwrz-ssave-02,spsjt-sit-02} \\
\bottomrule
\end{tabular}
\end{center}
\end{table}

\begin{figure}[h!]
\centering
  \includegraphics[width=0.475\textwidth]{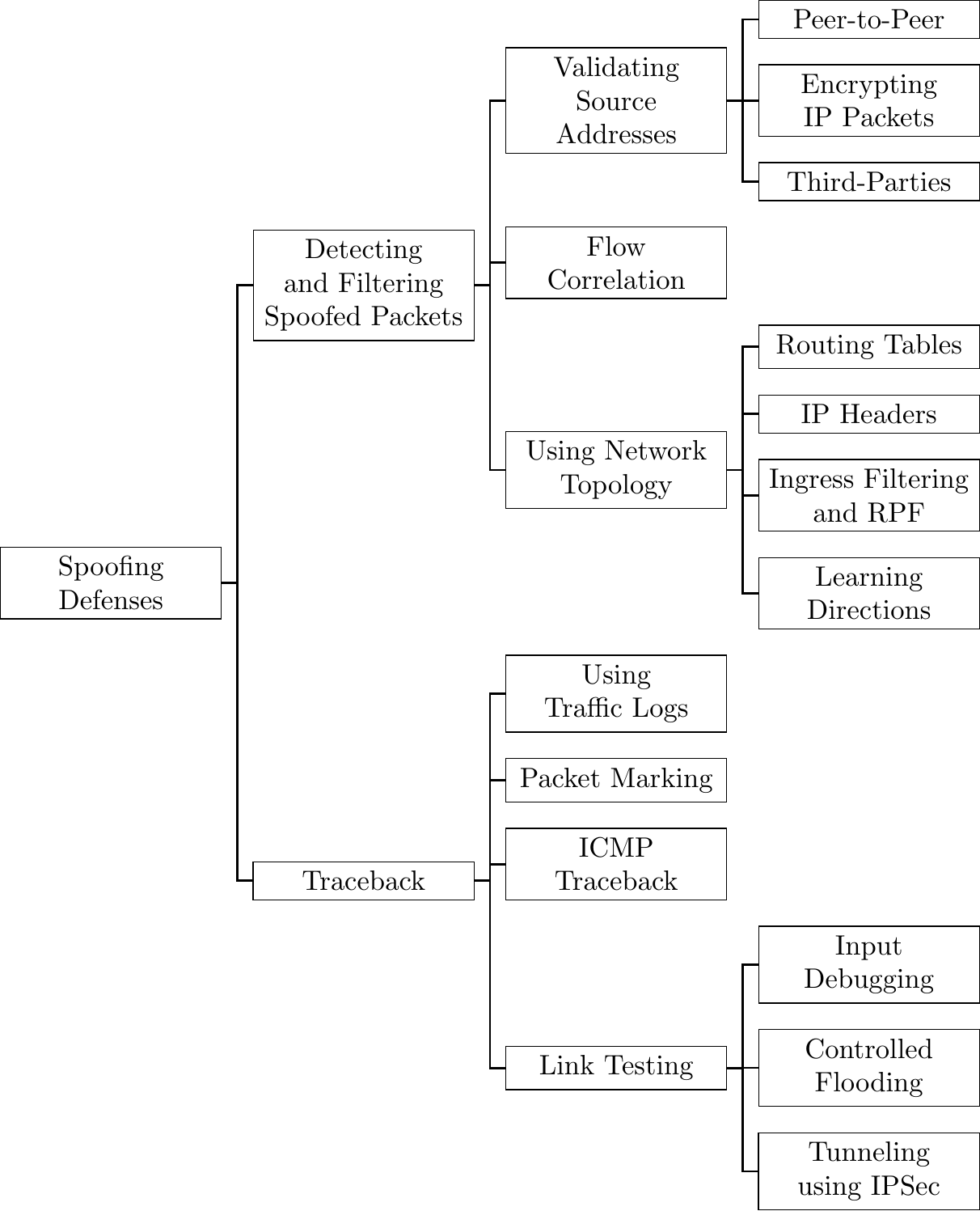}
  \caption{A categorization of source IP address spoofing defenses.}
  \label{fig:spoofingcats}
\end{figure}

We will start this section by discussing the techniques used to monitor spoofing and to map networks which do not guard against it. Later we will (a) survey the contributions listed in Table~\ref{table:surveys} in a single concise discussion, and (b) significantly extending the discussion to include new approaches. This will involve categorizing spoofing defenses as being for either packet filtering or traceback, as shown in Figure~\ref{fig:spoofingcats}.

\subsection{Monitoring Spoofing In The Wild}

There are two techniques which are being used to look at spoofing on the Internet. \emph{Active monitoring} attempts to send spoofed packets and monitor the replies, while \emph{passive monitoring} requires analyses of Internet traffic, e.g, NetFlow data. In the following two subsections we examples of both approaches.

\subsubsection{Active Monitoring}

The MIT Spoofer Project~\cite{bb-spies-05} measures how susceptible the Internet is to IP spoofing. To achieve a high coverage they have created and distributed an application which volunteers can install. The application sends several spoofed UDP packets to a server controlled by the project. The application then connects to the server using legitimate means to determine which packets were received or lost. The results of the MIT Spoofer Project have shown that about 25\% of ASs are currently vulnerable to IP spoofing. Daily statistics are also available on their website~\cite{b-spsis-14}.




K\"uhrer et al.~\cite{khrh-ehria-14} leverage the fact that some DNS proxies do
not correctly change the IP addresses when forwarding DNS packets to test if their
AS allows IP spoofing. They found 2,063 ASes that allow spoofed traffic by detecting if DNS replies have a different source IP addresses than the destination of the initial request.

The disadvantage of their approach is that it relies on networks which have open DNS resolvers that do not change the source IP address on responses. However, their approach does not depend on the distribution of volunteers like the Spoofer Project, and the researchers were largely free to choose which networks to investigate due to the many available open DNS resolvers. 

\subsubsection{Passive Monitoring}

An orthogonal question to see who is spoofing, is to measure how much Internet traffic is actually being spoofed. The UCSD Network Telescope from CAIDA~\cite{msbvs-ida-06} is attempting to answer this question by monitoring traffic which is sent to darknets. A darknet is an address-space which is routable but where no traffic is expected to be sent, i.e., the addresses are not allocated. Monitoring traffic to darknets can be used to spot attacks, scans of the Internet, and misconfiguration of machines. 

To explain how a darknet can be used to detect spoofed packets we will briefly describe the steps of a TCP-SYN flooding attack~\cite{rfc4987}. When an attacker is using spoofing for a TCP-SYN flooding attack and randomly chooses a source address for the spoofed packets, there is a chance that the address is in the address space of the darknet. As a consequence, the victim of the attack will response to the spoofed addresses, and the SYN/ACK packets from the victim will appear as traffic to the darknet. These SYN/ACK packets are known as backscatter, and since the darknets are supposed to cause no traffic, backscatter is assumed to be the result of spoofing once misconfiguration has been ruled out.

\subsection{Validating the Source IP Address}

There is no built in mechanism in the Internet Protocol which can prevent spoofing. Its function is simply to provide best effort delivery of datagrams~\cite{rfc791}. Nevertheless, over the next two subsections we will discuss various techniques which can be used to filter spoofed packets. In this section we will concentrate on methods which use some form of authentication to validate packets as they arrive and filter those which are not validated. In the next section we will look at more probabilistic methods.

\subsubsection{End-to-End Address Verification}

There are many end-to-end methods with which peers can attempt to validate the source IP address of packets. For example, probes can be used to provoke responses from the source address to see if it sent the packet~\cite{kc-mmtsf-11}. However, packet loss means that some probes may be lost, and other unpredictable behaviour in the Internet means that probing cannot guarantee that the packet was spoofed.

Additional information embedded in packets and which can be validated using a peer-to-peer network has been used to authenticate the source IP address~\cite{llyxw-psasa-08, llzc-memei-08, bl-spm-05, ls-tspbh-07, vp-drist-12, b-sc-96}. This information can be unique between two peers, and may also have a time limit associated with it. The literature also uses different terminology to refer this data. Common notations are ``cookie''~\cite{b-sc-96,ffl-dinp-05,gcc-sdpda-06} or ``key'' if the data is used in cryptographic functions~\cite{bl-spm-05,llyxw-psasa-08,kg-ssavh-09}.

Whilst the basic principle of using end-to-end methods to validate the source IP address is the same, all of the proposals have their own strengths and weaknesses. For example, large ``puzzles'' sent by servers may themselves be used in amplification attacks~\cite{ffl-dinp-05}, and there are additional processing and space overheads associated with generating and stamping ``passports'' that are appended to packets~\cite{llyxw-psasa-08}. Some of these overheads can be lowered, for example, if only routers at the edge of networks add data to, and validate, packets. However, this implies cooperation between network operators to install new functionality at the edges of their networks when the benefits to them may not outweigh the extra costs.


Lu et al.~\cite{llzc-memei-08} proposed Multi-purpose Anti-Spoofing Key (MASK) where packets are tagged with MASK labels which are unique between the source and destination routers. At first, MASK labels are exchanged using a TCP connection. Then, MASK labels can be included in subsequent packets by overwriting the ID field in the IP header. However, MASK needs to be evaluated in more realistic scenarios. Overwriting the ID field means that datagrams can not be fragmented. It is also not clear how big MASK labels are, and if they will fit in the 16-bit ID field. Furthermore, MASK appears to be susceptible to eavesdropping attacks as MASK labels are not exchanged securely and are also sent in plain text in subsequent packets.

Another example is the Source Address Validation Architecture with Host Identity Protocol (SAVAH)~\cite{kg-ssavh-09}, which can filter spoofed packets at the edge of local networks. It requires that all end-hosts and gateway routers support the Host Identity Protocol (HIP)~\cite{g-hipts-08, rfc4423}. SAVAH filters spoofed packets by requiring senders replace the source IP address of packets with a hash. The hash is created using the contents of each packet and a key agreed previously when the sender joined the network. This is cryptographically stronger than \cite{llzc-memei-08} so long as the key was exchange securely. However, this approach degrades performance of the network rapidly due to the processing overheads of hashing every packet. 

\subsubsection{Encryption}

An overall more secure approach than appending validation data to packets is to encrypt all IP packets. Using encryption between peers would not only prevent most spoofing attacks, but it also has the added benefit of all Internet traffic being secure by default~\cite{rfc4301}. One suitable mechanism by Gilad and Herzberg provides relatively lightweight encrypted communication~\cite{gh-ldais-12}.

However, the computational overheads and extra traffic caused by encrypting all IP traffic is currently impractical, and would also conflict with IoT scenarios using constrained devices. Nevertheless, encryption can be used in select cases such as communication with amplifiers. 

\subsubsection{Third-party Authentication}

Sometimes it is not desirable for routers and end-points to use end-to-end address verification, or negotiate keys and encrypt packets because of the additional processing and latency overheads. Using extra third-parties which are usually not involved in packet forwarding offers a way of authenticating packets which can potentially lower the extra processing demands placed on routers and end-points~\cite{iyam-pmaua-01, gaj-taaia-11, nh-bspsm-09, nh-bspsm-12}. 

For example, Gonzalez et al.~\cite{gaj-taaia-11} proposed that routers should send copies of packets to a ``judge'' (the trusted third-party). The judge should know the IP addresses serviced by the routers in order to make a decision whether or not the packet is spoofed, and the judge can also query routers in the absence of data. 

Noureldien et al.~\cite{nh-bspsm-09, nh-bspsm-12} suggested that authentication servers in local networks can be used to determine if TCP-SYN packets are spoofed. They propose that all packets that exit a network are inspected by an authentication server. An authentication server can also check incoming packets with the authentication server at the source network. This approach can lower processing overheads for some hosts in the network but it does not address concerns about additional latency and traffic volume.




\subsection{Flow Correlation}

Flow telemetry can help to identify anomalies in network traffic by inspecting statistical information of traffic that is aggregated by source and destination addresses/ports.
Kreibich~et al.~proposed to inspect asymmetries in packet counts per flow direction to find reflection attacks from the victim's perspective~\cite{Kreibich2005}.
Rossow proposed to leverage volume asymmetries to identify amplification attacks, both from the amplifier's and victim's perspective~\cite{r-ahrnp-14}.
Related methods that target non-amplification attacks aim to detect TCP-SYN flooding attacks by observing unbalanced SYN and SYN-ACK packet flows~\cite{spslj-hit-01,spsjt-sit-02,cy-datsf-06,kgsd-spita-07,njh-dsfau-08,sxll-lithi-08}. 


\subsection{Filtering Spoofed Packets Using Network Topology}

Almost all of the filtering methods seen so far come with a high cost in terms of additional latency, traffic volume, and/or processing at routers. In this section we will discuss methods which can lower the overheads by using what information is already known about the structure of the Internet. For example, some IP addresses are not meant to be used outside of local networks. Packets with private IP addresses (e.g., 192.168.1.1) should not be routed, as Network Address Translation (NAT)~\cite{rfc1631} at the source network should have changed the source address of the packet.

More generic methods of detecting spoofed packets, as discussed next, are probabilistic because the Internet is both dynamic and complex. However, some of these approaches also include a mechanism to determine if a packet was actually spoofed following detection.

\subsubsection{Using IP Headers}
\label{sub:ipheadersforspoofing}

One way with which spoofed packets can be filtered without using additional bandwidth is to analyze information present in the IP headers. If the legitimate values for a source IP address is known, then spoofed packets which do not have the correct information can be identified and filtered.

For example, spoofed packets can be filtered if their hop count does not match what is known about the topology of upstream routers. This information can be estimated using the Time To Live (TTL) field in the IP header~\cite{jws-hfeda-03, wjs-dasit-07,azh-pfbsr-07}. However, there may be many paths between the source and destination of spoofed packets. One interesting idea is to use Ant Colony Optimisation (ACO) algorithms to find these paths~\cite{av-dmdis-12}, and to collect TTL values which can be used to filter spoofed packets.

It is even possible to filter spoofed packets by detecting what operating system a remote host is using, i.e., by estimating the initial TTL of packets~\cite{t-arpof-03,f-rosd-06}. However, filtering spoofed packets based on the TTL and other fields does not guard against spoofing attacks where the attacker can manipulate the values in the IP header.



\subsubsection{Ingress Filtering and RPF}

Filtering IP packets with spoofed addresses can be achieved using knowledge of the IP addresses allocated by the upstream or downstream networks~\cite{rfc3013, rfc2827, rfc2267}. \emph{Ingress} filtering filters incoming packets. In contrast, \emph{egress} filtering filters packets which are exiting the network.

Ingress filtering (BCP 38) is sometimes implemented at the periphery of the Internet to stop packets with spoofed addresses being routed by edge routers~\cite{rfc2827}. However, not all network operators implement BCP 38. Ingress Access Lists are also typically maintained manually, and having outdated or misconfigured lists can prevent legitimate or allow spoofed traffic~\cite{rfc3704}.

Unicast Reverse Path Forwarding (uRPF) is an implementation of BCP 38 and uses forward routing information to filter spoofed packets~\cite{cs-urpfe-05}. Incoming packets are checked against a routers Forwarding Information Base (FIB) to ensure that packets are only forwarded if they come from the interface which is on the router's best path to the source address. uRPF also has a ``loose mode'' where it can check the source addresses packets without taking the interface into account. This allowed uRPF to be used on routers with multiple links to multiple ISPs~\cite{cs-urpfe-05}.

\subsubsection{Learning Directions}
\label{sub:LearningDirections}

Routers can record information from incoming packets in order to filter traffic which comes from unexpected directions~\cite{mjr-pisdt-06, el-idplv-07, wrl-savap-07, lmewr-lvidi-08, sgd-pfsv-08,onsw-peidf-07}. The Source Address Validity Enforcement (SAVE) protocol~\cite{lmwrz-ssave-02} is one example where routers can learn the expected incoming interface for a source address in order to authenticate subsequent packets.

The approach proposed by Wu et al.~\cite{wrl-savap-07} is called Source Address Validation Architecture (SAVA). To prevent IP spoofing in a local network, SAVA routers bind the source IP address and MAC address to a specific switch port. The next step is to filter packets at the Intra-AS level. This is done by associating source addresses with the incoming interface. The last step, filtering spoofed packets at inter-AS level, involves adjacent ASes exchanging address blocks so that packets from other ASes can be filtered.

However, these methods do not perform well during legitimate changes to the Internet's structure or when there are multiple paths between two routers~\cite{mjr-pisdt-06}. To counter this problem these methods can consider multiple valid interfaces per source address at the cost of lower filtering accuracy. Mirkovic et al.~\cite{mjr-pisdt-06} proposed dropping some TCP packets which originate from new interfaces. Under normal circumstances these packets should be retransmitted and the operation of the Internet is not severely compromised. However, this approach does not help with combating amplification attacks which are perpetrated using UDP or where the attacker retransmits attack packets.

\subsubsection{BGP and Routing Tables}

BGP is responsible exchanging routing information between ASes. As a result, many papers have proposed that BGP can also be used to help filter spoofed packets~\cite{pl-erpfd-01, dyc-cipfc-06, dyc-cipti-08}, e.g., by adding anti-spoofing information to BGP update messages~\cite{lkhp-bidmv-07} using packet marking techniques such as Pi which we talk about in Section~\ref{sub:marking}.

One way to use BGP routing tables to filter spoofed packets is to use Distributed Packet Filtering (DPF)~\cite{pl-erpfd-01}, and its extensions Inter-Domain Packet Filters (IDPF)~\cite{dyc-cipfc-06, dyc-cipti-08} and Extended IDPF~\cite{vp-drist-12}. Using IDPF a router examines BGP routing tables to check if it is on the optimal path between the packet's source and destination. However, this approach suffers from the same security flaws as BGP because there is no built-in mechanism to check the integrity and source of BGP messages~\cite{bfmr-baabg-08}. To improve the security of BGP and thus IDPF, Dawakhar et al. have suggested using Credible BGP~\cite{igm-misvb-09}.

\subsection{Tracing the Attacker}
\label{sub:ipspoofingtraceback}

Next to detecting and filtering spoofed IP packets, an orthogonal challenge is to trace the actual source of spoofed traffic~\cite{ba-oit-03}. Such tracing is useful, not only to stop attacks as they are happening by filtering the attack traffic at the source (thus saving the resources at upstream networks), but also so that the perpetrators can be prosecuted and further attacks prevented.

Tracing spoofed packets is difficult because of limited access to external routers and the the high overheads involved. The existing traceback vary in their objective and where they are applied. For example, an administrator of a local network might want to know which MAC address or interface is being used to send spoofed packets~\cite{ab-asdpa-07}, whereas an attack victim might want to know which network the attack originated from.


In amplification attacks, the victim receives attack traffic from amplifiers.
Thus, the traffic facing the victim is not spoofed.
Instead, spoofing detection and tracebacks needs to be performed at the amplifiers, possibly with the help of amplification honeypots~\cite{kkmnk-amdad-15}.

%

It is also important to realize that the traceback can only be used to identify the source of spoofed packets. Detecting the \emph{individuals} involved requires many out-band techniques such as monitoring the chat networks used by cyber criminals.
Furthermore, tracing becomes inherently difficult in case of distributed attack sources, such as DDoS botnets.


\subsubsection{Using Traffic Logs}
\label{sub:trafflogs}

Tracing spoofed packets back to the attacker can be done by analyzing traffic flow data collected by routers~\cite{spsjt-sit-02, spslj-hit-01, sxll-lithi-08}. With log-based traceback there is always the potential to trace a single packet~\cite{kgsd-spita-07}, and to trace attackers who use reflection. However, these methods often require that traffic data is either collected or stored by routers~\cite{s-ciont-00}. This comes with many practical difficulties such as additional hardware costs, as was the case with the original proposal by Snoeren et al.~\cite{spsjt-sit-02}, which requires Source Path Isolation Engine (SPIE) routers in place of existing routers. One solution to replacing existing routers is to use tap devices which eavesdrop on traffic~\cite{sjtss-tsips-03}. Finally, one can lower the amount of storage required to trace attackers by using hashes~\cite{spslj-hit-01}.

\subsubsection{Packet Marking}
\label{sub:marking}


The idea of packet marking is to record the path which packets follow into the packets~\cite{swka-pnsit-00}. Early packet marking schemes appended the IP addresses of routers into packets~\cite{swka-pnsit-00,dkk-ursis-00}. However, the four addition bytes needed to represent an IPv4 address and the large number of hops quickly adds to the size of packets. Song and Perrig suggested reducing the amount of space required by encoding addresses using hashing and a map of upstream routers~\cite{sp-aamsi-01}. Yaar et al.~discussed ways to lower the space needed by using shorter identifiers than IP addresses~\cite{yps-ppimd-03}. However, appending data unnecessarily to packets is still undesirable because it can increase fragmentation. To address this problem some packet marking schemes also suggested overloading IP header fields~\cite{swka-pnsit-00,sp-aamsi-01,dfs-aait-02,yps-snpmf-06,cpm-dstdd-07} and only encoding the addresses of edge routers~\cite{ga-prims-07}. 

Savage et al.~introduced Probabilistic Packet Marking (PPM) to reduce size and processing overheads of packet marking even further~\cite{swka-pnsit-00,swka-nsit-01}%
\footnote{Shokri et al.~\cite{svmy-dddpm-06} had a similar idea 6 years later and called it Dynamic Marking.}. PPM can be used to identify network path(s) traversed during an attack so long as the victim receives sufficient marked packets. Dong et al.~\cite{dabh-eppm-05} designed a single-bit-per-packet scheme. However, Adler~\cite{a-tppmi-05} noted that there is a trade-off between the number of bits allocated to PPM and the number of packets that must be received by the victim. 

Duwairi et al~\cite{dcm-eppms-04} proposed a hybrid PPM and data storage method called Distributed Linked-List Traceback (DLLT). The ``store, mark, and forward'' operation used by DLLT requires a ``marking field'' in each packet big enough to store a single IP address, and a data structure on each router in which to store information about forwarded packets. Any router which marks a packet must first store the address in the marking field (if there is one) along with a packet identifier, then the router is free to write its address in the marking field before forwarding the packet. A linked list is inherently created using this operation because the address in the marking field points to the previous router, which has a record of where it received the packet from, and so on.
However, its difficult to trace attacks with multiple attack sources~\cite{pl-eppmi-01}. To counter this, Song and Perrig looked at the security of markings and proposed using time-released key chains to authenticate them~\cite{sp-aamsi-01}.

Li et al.~\cite{lwys-tda-11} proposed an efficient packet marking scheme specifically for reflection attacks called Authenticated Deterministic Packet Marking (ADPM), which is an extension of the approach proposed by Belenky et al.~\cite{ba-dpm-07}. ADPM assumes that all routers are capable of matching request and response packets and that routers adjacent to reflectors cooperate to copy packet markings from the request packets to the replies. 


Dean et al.~phrase packet marking as a polynomial reconstruction problem and encodes path information as points on polynomials~\cite{dfs-aait-02}. Chen and Lee also extended their work to target reflection attacks~\cite{cl-aiptta-03}. They did this by requiring reflectors copy packet markings from the request packets to the replies. This enables the victim to trace paths to the attacker using less than 4,100 packets, but it may be possible to combine logging at routers and PPM to achieve traceback with a smaller number of marked packets~\cite{ag-nhsep-06}

A drawback of most packet marking schemes is that they rely on routers to maintain and update the markers.
In case routers drop or do not extend the marks, packet marking becomes significantly less effective.

\subsubsection{ICMP Traceback}

Bellovin et al.~proposed iTrace where routers send ICMP messages to the destination address of packets they forward so that the destination can see the path which the packet followed~\cite{bl-itm-00}. ICMP traceback can also be probabilistic in that the chance of a router generating an ICMP message is small, thus lowering overheads in a similar way to PPM~\cite{mmwwz-deidi-01,ltxm-itcpe-03, lkc-ppbit-04}. 
Mankin et al.~improved iTrace with ``intention-driven'' traceback after they noticed that the chance of a router picking an attack packet close to the attacker is much smaller than closer to the victim for certain DoS attacks~\cite{mmwwz-deidi-01}. Wang and Schulzrinne also proposed a new ICMP message which can be used in ICMP traceback for reflection attacks~\cite{ws-aiptm-04}.

\subsubsection{Link Testing}

Link testing is the systematic testing of intermediary links between routers, such as implemented via \emph{input debugging}~\cite{s-ciont-00}, DECIDUOUS~\cite{cnwvb-ddsin-99}, or \emph{controlled flooding}~\cite{bc-tapta-00}, as described next.

Input debugging attempts to determine from which upstream router an attack is coming from by recursively generating attack signatures to distinguish malicious packets from normal traffic. However, attack signatures that only match malicious traffic are difficult to create. Good attack signatures match only attack traffic and a only little to no legitimate traffic~\cite{s-ciont-00}.

The Intruder Detection and Isolation Protocol (IDIP)~\cite{sds-iidr-2000} describes an architecture which also uses attack signatures and can be used for link testing and attack mitigation. In IDIP a single Discovery Coordinator sends trace messages to other IDIP enabled devices. The trace messages contain a description of the event, i.e., an attack signature, using the Common Intrusion Specification Language (CISL)~\cite{fkpss-cisl-99}. IDIP devices which receive trace messages reply with report packets. The Discovery Coordinator can then decide what action to take, including instructing other IDIP devices to block attack traffic for a short amount of time~\cite{sds-iidr-2000}.

Another link testing method uses IPSec~\cite{RFC-1825,RFC-4302,RFC-4303} to trace the network from where a spoofed packet originated. For example, the DECIDUOUS implementation~\cite{cnwvb-ddsin-99} creates secure tunnels between itself and upstream routers until it finds the network from where the attack is coming from. However, DECIDUOUS needs to have a complete overview of the network topology to choose routers with which to tunnel. Furthermore, DECIDUOUS has high processing and traffic overheads caused by setting up and tearing down secure tunnels.

Controlled Flooding is a link testing method which floods each upstream router with packets. This technique needs the knowledge of the topology of the Internet. Thus, a map of the Internet is obtained before link testing~\cite{bc-tapta-00}. The purpose of flooding upstreaming routers is to determine if the attack traffic affected. If the attack traffic is disturbed significantly when sending traffic to a router then it is assumed that this particular router is one forwarding attack traffic.


\subsection{Discussion}

We have surveyed approaches to defend against spoofing. However, there are also more theoretical works in this area. For example, given a graph of the network topology and a set of firewall rules for each node, Santiraveewan and Permpoontanalarp proposed a methodology which can check to see if spoofing is possible~\cite{sp-gmais-04}. 

With all of these defenses, it may be confusing as to why spoofing is still a problem. There are many practical and economic reasons. Many of the proposals incur extra latency, traffic volume, computation, and/or storage overheads. Others are expensive to implement, as they require new hardware or staff to maintain them. Finally, it is not clear how well the methods will perform with only limited deployment~\cite{pl-erpfd-01}, and what rewards they offer to early adopters~\cite{lkhp-bidmv-07}.

We have seen that it is to difficult to prevent spoofing or trace the attackers in reflection/amplification attacks without considerable investment and cooperation by the different network operators. Therefore, we have to look for other ways of detecting and mitigating amplification attacks as soon as possible. The next section will address this topic in more detail.

%% file: DRDoS_Detection.tex
\section{Detecting and Mitigating Amplification Attacks}
\label{sec:reacProt}


The preventative measures described in Sections~\ref{sec:prevProt} and~\ref{sec:ipSpoofing} are effective defences against amplification attacks. Nevertheless, we have seen that they cannot always be relied upon. So the defences described in this section are to be used when preventative measures fail.

Defending against amplification attacks involves \emph{attack detection} and \emph{attack mitigation}, e.g. packet filtering or black holing. Attack detection is easier to perform near to the victim because traffic flows resemble a funnel with many attackers sending packets to a single location~\cite{zjt-asdma-13}. In contrast, attack mitigation is best performed closer to the source to save the resources of the victim and upstream networks.



Chang~\cite{c-dafdd-02} explained that DoS defenses can be located at four areas: (i) the victim's network, (ii) the victim's ISP network, (iii) further upstream networks, and (iv) attack source networks. 
Zargar et al.~\cite{zjt-asdma-13} classifies DoS defenses into two categories: (i) network/transport-level, and (ii) application-level. They also sub-categorized DoS defenses based on their deployment location: (i) source, (ii) destination, (iii) network, and (iv) distributed/hybrid. 

\begin{figure}[h]
\centering
\includegraphics[width=0.475\textwidth]{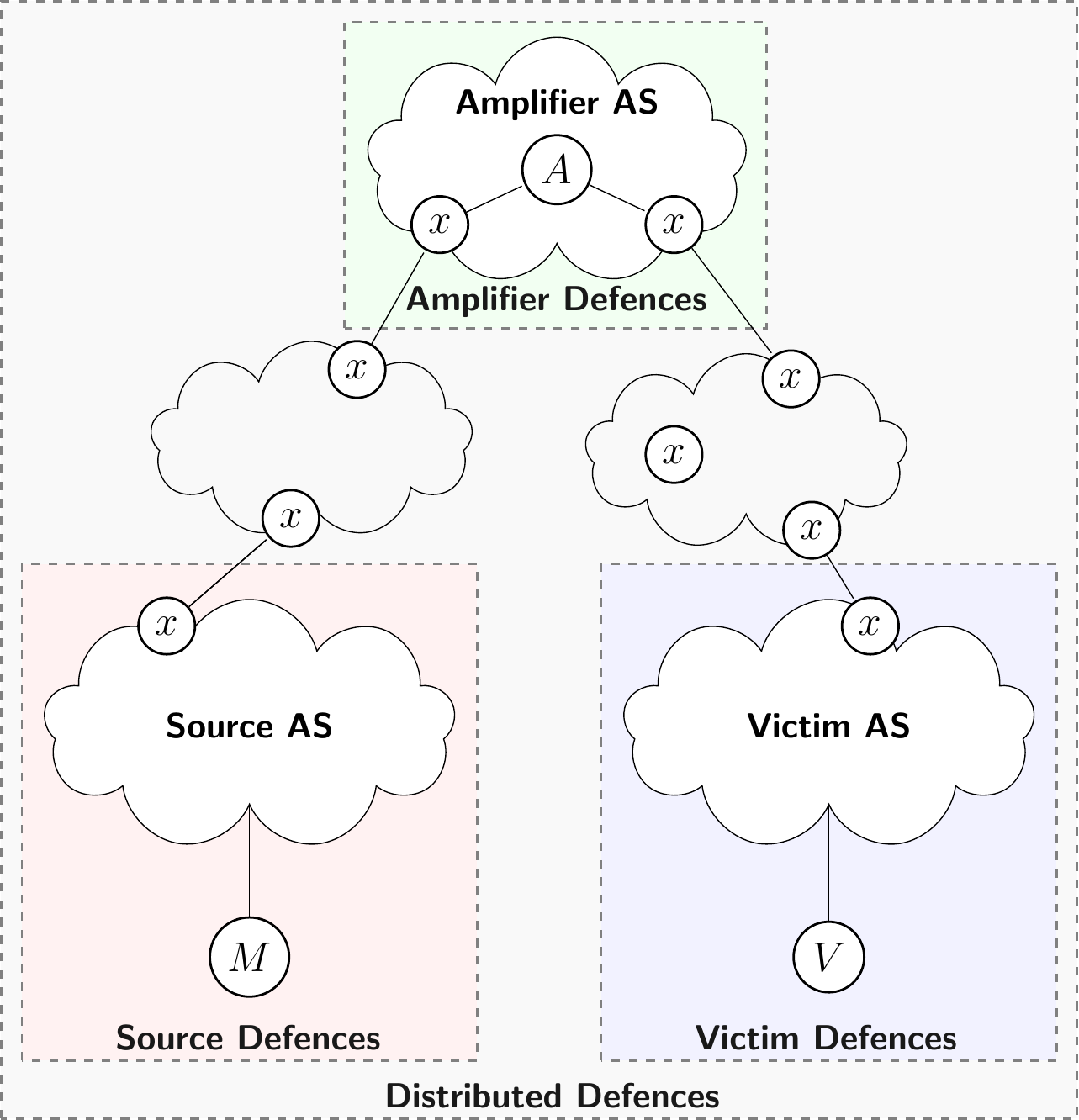}
\caption{Locations for defending against amplification attacks. Routers are marked $x$.}
\label{fig:defenselocations}
\end{figure}

However, in this survey we are concerned specifically with amplification attacks, and we observed that it is also possible to defend against amplification attacks at the amplifiers. Therefore, we have have split amplification attack defenses into the following location categories described in Figure~\ref{fig:defenselocations}: (i) at the victim's network, (ii) at the source network(s), (iii) a single unspecified location in cases where the authors have not been clear about the deployment location, (iv) distributed, and (v) at the amplifier(s). In cases where the amplifier itself is the target of an amplification attack, defenses will be limited to source-based and amplifier-based defenses.

It is also important to note that not all of the DoS defenses surveyed by Chang and Zargar et al. are applicable to amplification attacks. The most devastating attacks we described in Section~\ref{sec:amplOver} originated from multiple sources, and combined both traffic amplification and reflection. Therefore, in this section, we will discuss the defenses in Table~\ref{table:detectionMethods} which include new proposals that specifically target amplification attacks, but we will also include some older proposals which can part-defend against amplification attacks, e.g., defenses against reflection.



\begin{table*}
\centering
\caption{Methods for detecting and predicting amplification attacks. Some of these proposals also include novel techniques which aim to identify and filter attack packets.}
\label{table:detectionMethods}
\begin{tabular}{ p{5.25cm}p{7.5cm}p{1.5cm}cc }
\toprule
\textbf{Method and References} & \textbf{Description} & \textbf{Location} & \multicolumn{2}{c}{\textbf{Time}} \\ 
 &  & &  \textbf{Before} & \textbf{During} \\ 
\midrule
Attack pattern correlation~\cite{wcxj-rcdba-13} & Tries to detect reflection attacks by correlating the arrival rate of packets at the victim with known attacks. & Victim & & \checkmark  \\
 \midrule
Missing request packets~\cite{kmgg-fsdaa-07, kmgg-ddaa-08,dl-padra-11,tnoy-srpcm-06,toyaw-ddasr-08} & Attempt to match incoming responses to previously sent requests. & Victim & & \checkmark \\
 \midrule
History-based IP Filtering (HIP)~\cite{plr-pddos-03} & Attempts to filter packets during periods of high congestion which come from previously unseen addresses. & Victim & & \checkmark \\
\midrule
Comparing incoming and outgoing packets at source networks (D-WARD)~\cite{mpr-ads-02, mpr-sddod-03} & Uses protocol specific knowledge to predict the number of expected responses, e.g., replies to ICMP requests. Can also rate limit attacks at the source. & Source & & \checkmark  \\
 \midrule
Comparing the volume of incoming and outgoing traffic~\cite{gp-mdbad-01,aglrs-efdos-03} & MULTOPS~\cite{gp-mdbad-01} and TOPS~\cite{aglrs-efdos-03}  & Unspecified & & \checkmark \\
\midrule
Detection with Information Sharing (DIS)~\cite{plk-drasb-03} & Aimed at TCP based reflection attacks, but reflectors collaborate to improve the accuracy of attack detection. Similar collaboration could potentially improve attack detection at amplifiers. & Reflectors / Amplifiers & & \checkmark \\
\midrule
PacketScore~\cite{klcc-psspf-06} & Routers score packets, and communicate with a central server for a global view and filter packets with a high score. & Distributed & & \checkmark \\
\midrule
Repetitive packet sizes~\cite{h-dradf-13} & Monitoring DNS packet sizes as well as the number of packets across multiple routers.  Suspicion is based on thresholds calculated using real data from the GEANT network. & Distributed & & \checkmark  \\ 
 \midrule
Comparing payload bytes of request and response packets~\cite{r-ahrnp-14} & Analysed NetFlow data to detect pair-flows where the response payloads are larger than the request payloads (a similar idea to that which is used in  MULTOPS and TOPS). Limited to 14 UDP protocols. & Distributed & & \checkmark  \\
 \midrule
Comparing payload bytes of request and response packets, plus compare packet sizes of requests~\cite{bbger-doaap-15} & Combines techniques from Rossow~\cite{r-ahrnp-14} (comparing request and response sizes) and Huistra~\cite{h-dradf-13} (similar packet sizes between requests). & Distributed & & \checkmark \\
 \midrule
Aggregate-based Congestion Control (ACC)~\cite{mbfip-chban-02} & Looks for high bandwidth aggregates (collections of packets sharing the same destination address). Uses pushback (also used in \cite{ylly-dadda-05}) to instruct upstream routers to rate-limit flows. & Distributed & & \checkmark \\
 \midrule
Spoofing protection for amplifiers~\cite{gcc-sdpda-06} & Specific to amplification attacks. This method detects spoofed packets which do not have the correct session key provided by the amplifier for the source address. & Distributed & & \checkmark \\ 
\midrule
AD and PAD~\cite{cp-adtdd-05}, and TRACK~\cite{cpm-track-06} & A congested victim instructs upstream routers to use packet marking in order to trace the attack and then requests they filter the traffic. These methods may need to be updated for amplification attacks to take reflection into account. & Distributed & & \checkmark \\
\midrule
Game theory to predict attacks~\cite{fg-bvstm-09} & It is possible to study the motivation of attackers and the security of the Internet to predict attacks. & n/a & \checkmark \\
\midrule
Darknets to detect scans~\cite{fbd-tfmdd-13,fbd-fidnd-14,fbd-idrds-15,r-ahrnp-14} & Monitoring darknet traffic can detect random scans for amplifiers such as those seen before the Spamhaus attack~\cite{fbd-idrds-15}. & Distributed & \checkmark \\
\midrule
Honeypots to detect scans and attacks~\cite{r-ahrnp-14} & Honeypots can take the place of amplifiers and are a powerful way of detecting scans and amplification attacks. They also require careful configuration so they do not cause harm to the victim. & Amplifiers & \checkmark & \checkmark \\
\bottomrule
\end{tabular}
\end{table*}

\subsection{Victim-end Defenses}

Despite attack detection and packet filtering being separate objectives, many proposals attempt to do both near the victim.

Kambourakis et al.~\cite{kmgg-fsdaa-07, kmgg-ddaa-08} proposed a DNS Amplification Attacks Detector (DAAD) which logs outgoing DNS requests and incoming responses in lookup tables at the victim's network. Responses are matched to outgoing requests and responses that don't match requests are flagged as suspicious. The authors also propose that the DAAD should be able to alter firewall rules to filter packets. However, amplification attacks do not only misuse DNS as we described in Section~\ref{sec:amplOver}. So a more general detection method is desirable.


Tsunoda et al.~\cite{tnoy-srpcm-06} proposed a general method for matching request and response packets at a single point between the amplifier and the victim, e.g., at the victim's gateway router. They later extended their work with mathematical analysis and additional experimental results~\cite{toyaw-ddasr-08}. However, in order to correctly identify the responses for each request, a router needs to record the source and destination addresses, protocol field, application headers, etc. for every request it forwards. This requires memory-intense management operations. A partial solution to this problem is to use Bloom filters to efficiently store requests and check them against incoming response packets~\cite{dl-padra-11}.

History-based IP Filtering (HIP)~\cite{plr-pddos-03} attempts to filter packets during periods of high congestion which come from previously unseen addresses. It does this by adding addresses to a table during normal operation, and uses a sliding window to remove expired addresses. 

Wei et al.~\cite{wcxj-rcdba-13} proposed using traffic pattern correlation to look for patterns which are out of the ordinary, or which match known attack patterns. The advantages of using traffic pattern correlation and HIP for attack detection, are that they are lightweight and protocol independent. However, methods to detect attacks based on previous behavior can struggle to distinguish between attacks and unexpected bursts in legitimate traffic (flash crowds)~\cite{yzjgx-ddafc-12}. Furthermore, an attacker that knows these defenses can ``train'' the victim before the attack~\cite{plr-pddos-03}.

After an attack has been identified, traffic can be dropped before it enters the victim network by instrumenting the control plane.
A common approach is blackholing of IP prefixes or host addresses at Internet Exchange Points (IXP) \cite{dfk-biedm-16}.
In this case, a border router of the victim network signals a specific next hop for the addresses under attack, using BGP updates.
Traffic from neighboring routers to the victim will then be forwarded to this IP addresses (or more precisely to the MAC address), and finally discarded at IXP layer-2 ingress switches.

Detecting attacks at the victim's network may be ineffective because bandwidth may have already become saturated, or the volume of traffic is too much to process. This means that automated responses for filtering packets or alerting upstream routers to throttle traffic may not function correctly. In order to avoid this problem, another possible location for detecting and filtering amplification attacks is at the source network.

\subsection{Source-end Defenses}

Source-end defenses are desirable because they can stop attack traffic before it gets to upstream routers. However, source-end detection and filtering techniques may not be an effective defense against amplification attacks because: (i) the source of the attack can be distributed, meaning that the solutions need to be deployed on all networks, (ii) it can be difficult to differentiate between legitimate and attack traffic if each attacker only sends a few requests, and (iii) there is no immediate incentive for network providers to deploy source-end detection and filtering techniques because it is unclear what benefit it offers their customers.

One example of a scheme proposed to work at the attack source is by Mirkovic et al.~\cite{mpr-ads-02, mpr-sddod-03}. They proposed D-WARD specifically to detect DoS attacks which involve spoofed packets or high traffic volumes. They monitor both inbound and outbound traffic flows at the source network, and compare them with flow models derived from normal traffic. The authors also note that an attacker will not reduce its outbound traffic even when notified of congestion at the victim, so D-WARD can rate limit suspected source addresses at upstream routers.

\subsection{Distributed Defenses}

Victim-end and source-end defenses tend to be designed to run in isolation on a single machine, e.g., at a border router between networks. However, detecting and filtering amplification attacks at a single point in the Internet is problematic because of asymmetric routing, and the fact that attacks are distributed. This means that in order for isolated defenses to be effective, all inbound and outbound paths in the Internet should be symmetrical, and all networks should support the new defenses. 



Amplification attacks are distributed threats against many potential victims, and as a result they require distributed defenses.  Some distributed defenses also can detect and filter attacks after a victim has been taken offline by the attack and cannot use automated methods which use the affected connections to alert upstream routers~\cite{klcc-psspf-06,h-dradf-13,r-ahrnp-14,bbger-doaap-15}. It is even possible to detect and filter DoS attacks with limited deployment of a distributed protocol alongside legacy equipment/protocols~\cite{mrr-afdd-03}.

Kim et al.~\cite{klcc-psspf-06} proposed PacketScore which uses a DDoS Control Server (DCS) to collate reports from routers across the Internet. Routers with special reporting capabilities, called Detecting-Differentiating-Discarding Routers (3D-R), mark suspicious packets with a ``score'' which reflects how likely they are to be involved in an attack. The 3D-Rs then filter packets based on a variable score threshold provided by the DCS, which varies the threshold depending on the current load at the victim and the score distribution of attacking packets.

Mahajan et al.~proposed attack detection at victims in terms of ``aggregate'' flows, and proposed using pushback mechanisms to filter flows at upstream networks whenever a link experiences sustained severe congestion~\cite{mbfip-chban-02}. In pushback, the congested router asks upstream routers which are involved in the flow aggregate to rate limit traffic flows~\cite{mbfip-chban-02,ylly-dadda-05}. 

Chen and Park~\cite{cp-adtdd-05} proposed an Attack Diagnostic (AD) system in which DoS attacks are detected near to the victim, and packet filtering is executed close to the attacker. AD also combines some familiar techniques; packet marking (Section~\ref{sub:marking}) is used for traceback, and an approach similar to pushback is used to alert the source networks.
Similarly, TRACK combines packing marking with packet filtering~\cite{cpm-track-06}. However, AD and Track are currently not suitable defenses for amplification attacks because their traceback methods will only trace back to the amplifier, but not to the actual attacker.

It is also possible to detect amplification attacks by analyzing traffic flow data collected by multiple routers. Usually these methods require data from locations between the source and the amplifier, as well as data from between the amplifier and the victim~\cite{h-dradf-13,r-ahrnp-14,bbger-doaap-15}. However, it is not required to have data from between the attacker and amplifier simply to detect DDoS attacks. Yu et al.~\cite{yzjgx-ddafc-12} proposed using flow correlation to detect DDoS attacks, but unlike the correlating method proposed by Wei et al.~\cite{wcxj-rcdba-13}, Yu's method can distinguish attacks from flash crowds. This was possible because it has access to data collected by multiple routers in a ``community network''.

Huistra showed that amplification attacks can be detected specifically by monitoring DNS packet sizes as well as the number of packets across multiple routers~\cite{h-dradf-13}. Huistra's method is split into two parts. The first (which can also be used to detect attackers) focuses on detecting IP addresses generating a suspicious number of similar sized DNS requests. Suspicion is based on thresholds calculated using real data from the GEANT network~\cite{h-dradf-13}. The second part can be used to detect the victim, and looks at the number of large DNS packets sent to a single IP address. 

Rossow's analysis and detection of amplification attacks is similar to Huistra's but is not limited to a single protocol~\cite{r-ahrnp-14}. Instead, he focused on 14 UDP protocols which allow amplification. By analyzing traffic samples taken from routers belonging to a single ISP, Rossow was able to detect 15 real life amplification attacks against multiple victims by comparing sent and received bytes.


Based on the work by Rossow, B\"ottger et al.~\cite{bbger-doaap-15} developed a protocol-agnostic approach to detect amplification attacks. They introduce the following detection criteria, which are independent of a specific network service, but base on the assumption that attackers reuse attack requests: (i) similar packet size among all requests (and responses), (ii) similar payload among all requests (and responses), (iii) an increased amount of ICMP unreachable messages, and (iv) incorrect TTL values. They found that the first two criteria are most suitable for amplification detection.



\subsection{Defenses At Reflectors/Amplifiers}

Rossow and Huistra also showed that it is possible to detect open DNS resolvers that have been abused in amplification attacks by analyzing traffic data collected by routers in a similar way as mentioned in the previous subsection, but from an amplifiers perspective~\cite{h-dradf-13,r-ahrnp-14}. 


Another approach for filtering spoofed packets, which was specifically designed to defend against amplification attacks, requires that session tokens are sent along with requests to amplifiers~\cite{gcc-sdpda-06}. This requires that the first time a request is received from a new address, the amplifier replies with a unique session token which the sender should include in all future requests. 

Including the same session token in all requests has low processing overheads but adds to the size of each packet. Furthermore, it is prone to eavesdropping if session tokens are not encrypted. If the attacker can eavesdrop on the victims Internet traffic then they can also agree new session tokens with amplifiers the victim has not yet contacted.

Peng et al.~\cite{plk-drasb-03} focused on TCP based reflection attacks and used mechanics which might also improve the accuracy of detection at amplifiers. Their method monitored the cause of packets instead of just counting them. In their proposal, reflectors monitor incoming TCP RST packets and monitor those which have a corresponding SYN/ACK state for the outgoing connection. RST packets indicate that the reflector received a spoofed SYN packet from an attacker and sent a SYN/ACK to the victim, at which point the victim has replied with a RST packet. Peng et al.~suggested that collaboration between reflectors can be used to improve the accuracy of attack detection. Once an attack has been detected, their system sends a warning to all other participating reflectors instructing them to stop the attack. 

\subsection{Defenses at Single Unspecified Locations}

Not all DoS attack defenses are designed with a particular location in mind, or the location was not explicitly specified. This is problematic because the challenges vary among locations.

MUlti-Level Tree for On-line Packet Statistics (MULTOPS)~\cite{gp-mdbad-01} is one example where a router somewhere on the Internet is used to compare the volume of incoming and outgoing traffic to look for DoS attacks. However, if MULTOPS is used on central routers which route packets for many different addresses, then the tree data structure it relies on will consume a significant amount of resources. In fact, MULTOPS itself can be targeted by memory exhaustion attack~\cite{gp-mdbad-01,aglrs-efdos-03}.

Tabulated On-line Packet Statistics (TOPS)~\cite{aglrs-efdos-03} improves on the accuracy of MULTOPS by taking into account the protocol being used, and also improves efficiency by using fixed length tables instead of trees. This makes TOPS suitable for use on busy links~\cite{zjt-asdma-13}.

Both MULTOPS and TOPS assume that legitimate incoming and outgoing traffic is proportional~\cite{aglrs-efdos-03}. However, this is not always the case, e.g., when streaming video. Attackers may also try to counteract MULTOPS and TOPS by trying to balance incoming and outgoing data volumes during an attack~\cite{zjt-asdma-13}. 


\subsection{Predicting Future Attacks}

It is desirable to know about DoS attacks before they happen. To this end, Fultz et al.~\cite{fg-bvstm-09} proposed using a game theoretical approach to predict DDoS attacks. They found that attackers only launch attacks if defenders have not invested in adequate protection or if penalties, e.g, monetary costs and risks of being caught are low.

It is also possible to detect scans for amplifiers which may act as a warning for future attacks. For example, darknets (also known as blackholes) are unused IP address spaces on the Internet. Since the IP addresses in darknets are supposed to be unused, any traffic to or from a darknet is a sign of either scanning, misconfiguration, or malicious intent~\cite{ckgpb-tt8pg-14}. 


Fachkha et al.~\cite{fbd-tfmdd-13,fbd-fidnd-14,fbd-idrds-15} looked at 1.44TB of traffic data for a /13 address block of darknet IPs~\footnote{The data was provided to Fachkha et al. by Farsight Security https://www.farsightsecurity.com/.}, and found 134 separate incidents which they claimed may have been amplification attacks. 

However, Fachkha et al. may have detected scans for open DNS resolvers rather than actual attacks. To explore this idea we can look at Tables 7 and 8 in~\cite{fbd-idrds-15} to see the incidents they detected. Specifically, they detected 2 large events (February $19^{th}$ and March $18^{th}$) with very high packets per minute. March $18^{th}$ is interesting because it is the same day as the Spamhaus attack~\cite{p-dtbi-13}. However, this looks like a horizontal IP scan because it targets 50,257 separate IPs with 1 packet each, which is indicative of a horizontal scan rather than an attack.

We can also take a closer look at the Spamhaus incident by using Figure 6 of~\cite{fbd-idrds-15}, which shows the packets per hour going to the Darknet monitored by Farsight Security:

\begin{itemize}
  \item The first spike we are interested in is at 337 hours, which is on the $14^{th}$ of March. This is shortly before the Spamhaus attack which was reported by CloudFlare as taking place on March $18^{th}$~\cite{p-dtbi-13}.
  \item The second interesting data spike at 385-409 hours is also before the attack.
  \item The graph for the $18^{th}$ of March (the day of the attack) is one of the quietest periods shown.
  \item The next spike they detect is at 517 hours, which is 3 days after the attack.
\end{itemize}

By referencing what we know about the Spamhaus attack~\cite{p-dtbi-13}, we suggest that the information presented by Fachkha et al.~\cite{fbd-idrds-15} shows horizontal scans for open DNS resolvers which may be related to the Spamhaus attack. Furthermore, attackers are unlikely to waste bandwidth sending speculative requests to unknown IPs when they can try to maximise the impact of an attack by only sending requests to known amplifiers. This supported by the quiet period observed by Fachkha et al. on March $18^{th}$.


Finally, to support defenses and monitor attacks, one can set up amplification honeypots that emulate protocols that are vulnerable to amplification abuse. Honeypots can be used to log scanning activity and report attacks that abuse them. In addition, Kr\"amer et al.~show that honeypots can be used to derive signatures of attack traffic, such as domains abused in DNS requests~\cite{kkmnk-amdad-15}, which is useful for detecting future attacks.

\subsection{Discussion}


Most of the defenses described in this section focus on amplification attack detection and filtering at the victim or by using distributed algorithms. Amplification attack defenses at the source networks are more desirable, but it is unlikely that defences specific to a single attack, e.g., amplification attacks, will be implemented in the short-term and on a large scale. 

One contribution of this paper is to survey the defenses currently available for amplifiers and to make some suggestions for future directions. We found that this area has had relatively little attention compared to defenses at routers, but by adopting one of the approaches detailed in Section~\ref{sec:prevProt} it may be possible for amplifiers to prevent amplification attacks from happening in the first place.

%% file: conclusion.tex
\section{Conclusions}
\label{sec:conclusion}

This paper discusses the current state of the art of research proposals for detecting, preventing, and tracing amplification attacks. As part of this, we have also surveyed defenses against source IP address spoofing, which is essential for amplification attacks. 

We have concluded that preventing source IP spoofing is the effective way of defending against amplification attacks. However, spoofed packets can still be sent from large parts of the Internet. So we need effective methods to detect and filter DoS attack packets as close to the source as possible. 

Another subject for future work is to assess the defenses we have covered against one another using a wider list of empirical criteria, e.g., monetary cost, accuracy, memory overheads, levels of inter-AS cooperation, etc. Zargar et al.~\cite{zjt-asdma-13} surveyed the metrics used to asses DoS defenses and discussed their categorization in some of these terms. However, a more detailed study on individual defenses is crucial because all of the proposals we surveyed come with their own strengths and weaknesses. Some of which may not be obvious, and it is important for system administrators to be able to weigh up their options in a concise and meaningful way.

Promising amplification attack defences appear to be distributed DoS detection and filtering algorithms, which can mitigate attacks even when the victim has gone off-line and is unable to take counter-measures itself. However, these methods require more collaboration between network providers than is currently the case~\cite{zjt-asdma-13}. 


More research is required to close the attack vectors which are being used in amplification attacks, e.g, closing open DNS resolvers, patching NTP servers, and promoting ingress filtering. A number of ongoing initiatives aim at tackling these issues, as summarized in Table~\ref{table:ongoing}. 

\begin{table}
 \caption{Initiatives to Identify Amplifiers}
 \label{table:ongoing}
\begin{tabular}{ p{2cm}p{6cm} }
\toprule
\textbf{Reference} & \textbf{Description} \\ 
\midrule
Open Resolver Scanning Project~\cite{shadowserver}  & Scans for open DNS resolvers and provide an automated system to notify affected networks. \\
\midrule
Open DNS Resolver Project~\cite{openresolver} & Scans for open resolvers and allows to query for such resolvers in a certain IP address range. \\
\midrule
Measurement Factory~\cite{measurementfactory} & Maintains a list of DNS servers that are known to serve as globally accessible open resolvers. \\
\midrule
Open NTP Project~\cite{ntfoundation}  & Scans for NTP servers in IPv4 which can be used in an amplification attack. \\
\bottomrule
\end{tabular}
\end{table}
